\documentclass[12pt,letterpaper,notitlepage]{article}
\pdfoutput=1

\usepackage{epsfig,amsmath,amssymb,cite,color,multirow,ulem}

\usepackage{graphicx,wasysym}

\usepackage{tabulary}
\newcolumntype{K}[1]{>{\centering\arraybackslash}p{#1}}

\makeatletter
\newcommand{\thickhline}{%
    \noalign {\ifnum 0=`}\fi \hrule height 1pt
    \futurelet \reserved@a \@xhline
}
\newcolumntype{"}{@{\hskip\tabcolsep\vrule width 1pt\hskip\tabcolsep}}
\makeatother

\newcommand{\gev}{\,\, \mathrm{GeV}}

\begin{document}

\begin{titlepage}

\begin{flushright}
TTK-16-32
\end{flushright}

\vspace{15pt}
\begin{center}
\LARGE Observing the Top Energy Asymmetry at the LHC
\end{center}

\vspace{0pt}
\begin{center}
{\large S.~Berge$^1$ and S.~Westhoff\,$^2$}\\
\vspace{30pt} {
$^1${\it Institut f\"ur Theoretische Physik, RWTH Aachen University, 52056 Aachen, Germany}\\
$^2${\it Institut f\"ur Theoretische Physik, Heidelberg University, 69120 Heidelberg, Germany}
   }
\end{center}

\vspace{10pt}
\begin{abstract}
\vspace{2pt} 
\noindent
The top-antitop energy asymmetry is a promising observable of the charge asymmetry in jet-associated top-quark pair production at the LHC. We present new predictions of the energy asymmetry in proton-proton collisions at 13~TeV, including QCD corrections at the next-to-leading perturbative order. The effect of QCD corrections on the observable is moderate. With suitable phase-space cuts, the asymmetry can be enhanced at the cost of reducing the cross section. For instance, for a cross section of $1\,\text{pb}$ after cuts, we predict an energy asymmetry of $-6.5^{\,+0.1}_{\,-0.2}\%$ at the next-to-leading order in QCD. We also investigate scale uncertainties and parton-shower effects, which partially cancel in the normalized asymmetry. Our results provide a sound basis for a measurement of the energy asymmetry at the LHC during run II.
\end{abstract}

\end{titlepage}

\clearpage

\section{Introduction}\label{sec:intro}
As the LHC is providing us with unprecedented high-energy data, top-quark physics has reached a mature level in both theory and experiment. We are able to test properties and interactions of the top-quark with high precision in inclusive cross sections, as well as in specific kinematic regimes through differential distributions. Probing the top-antitop charge asymmetry is a precision test of quantum chromodynamics (QCD). It bares information on strong interactions of heavy quarks that cannot be directly obtained from cross-section measurements~\cite{Kuhn:1998kw,Halzen:1987xd}. Phenomenologically, the charge asymmetry is often approached by comparing angular distributions of top- and antitop-quarks, commonly expressed in terms of rapidity differences~(for an overview of LHC observables, see~\cite{Bernreuther:2012sx,Aguilar-Saavedra:2014kpa}). Precise QCD predictions of the rapidity asymmetry at the Tevatron have been obtained at next-to-next-to-leading order (NNLO) in perturbation theory~\cite{Czakon:2014xsa} and from soft-gluon resummation~\cite{Almeida:2008ug,Ahrens:2011uf,Kidonakis:2015ona}.

At the LHC, the inclusive rapidity asymmetry in top-pair production is very small for two reasons: It is induced by partonic quark-antiquark interactions, which are buried under the large (charge-symmetric) gluon-gluon background. And it is suppressed in QCD, since it arises only at the perturbative next-to-leading order (NLO). With suitable phase-space cuts, the rapidity asymmetry can be enhanced in the forward region, where quark-antiquark interactions make up a bigger part of the total production rate~\cite{Kuhn:2011ri}. However, a significant enhancement of the asymmetry comes at the cost of reducing the data set to a kinematic region that is experimentally challenging to access. The ATLAS and CMS collaborations have measured the rapidity asymmetry at center-of-mass (CM) energies of 7 and 8 TeV~\cite{Aad:2016ove,Aad:2015noh,Khachatryan:2015oga,Khachatryan:2016ysn} and the same observable in the boosted-top regime at 8 TeV~\cite{Aad:2015lgx}. Currently, statistical and systematic uncertainties strongly limit the sensitivity to the charge asymmetry through rapidity differences. The ultimate sensitivity during the 13-TeV phase of the LHC will crucially depend on reducing the systematic uncertainties sufficiently to allow for an observation of the asymmetry.

Can we find an observable of the charge asymmetry that is sizeable in magnitude and better suited for a precise measurement? If the top-antitop pair is produced in association with a hard jet, the rapidity asymmetry appears already at the leading order (LO) in QCD, and it is indeed large and negative. But NLO corrections are sizeable and positive and reduce the asymmetry to a quantity no larger than in inclusive top-pair production~\cite{Dittmaier:2007wz,Dittmaier:2008uj,Melnikov:2010iu}.

Alternatively, the charge asymmetry can be observed in jet-associated top-pair production through top-antitop \textit{energy} differences~\cite{Berge:2013xsa}. This so-called energy asymmetry is induced at LO QCD from partonic quark-gluon interactions, which are much more abundant than quark-antiquark states in proton-proton collisions at high energies. The energy asymmetry is therefore less affected by the gluon-gluon background and larger than rapidity asymmetries. Since the definition of the energy asymmetry relies on the jet kinematics, it is important to know the impact of QCD corrections on the observable. In this work, we investigate fixed-order NLO corrections and parton-shower effects using Monte-Carlo methods. We will show that and why these corrections are moderate and provide improved predictions of the energy asymmetry for the LHC at 13 TeV collision energy.

Other observables of the top charge asymmetry at the LHC have been proposed in top-antitop production in association with an additional photon~\cite{Aguilar-Saavedra:2014vta} or $W$ boson~\cite{Maltoni:2014zpa}. Both of these processes probe the top charge asymmetry in the quark-antiquark channel. While the emission of an additional electroweak boson suppresses the gluon-gluon background, the sensitivity to the charge asymmetry is limited by the smallness of the cross sections and the net asymmetry. It is interesting to pursue these alternative approaches complementary to the energy asymmetry, which probes the quark-gluon channel in hadronic top-antitop-jet production.

In Section~\ref{sec:observables}, we review the definition of the energy asymmetry and a related optimized observable. Fixed-order NLO QCD corrections are discussed in Section~\ref{sec:nlo} from a qualitative viewpoint and in Section~\ref{sec:lhc} numerically for the 13-TeV LHC. Parton-shower effects from multi-parton emission are the topic of Section~\ref{sec:shower}. In Section~\ref{sec:kinematics}, we show how to enhance the energy asymmetry with kinematic cuts and present our final predictions for the LHC. We conclude in Section~\ref{sec:conclusions} with an outlook on the discovery prospects of the asymmetry.

\section{Observables of the charge asymmetry}\label{sec:observables}
The top-antitop energy asymmetry as an observable of the charge asymmetry was first introduced in Reference~\cite{Berge:2013xsa}. Here we briefly review the definition and properties of the energy asymmetry, as well as of an optimized observable~\cite{Alte:2014toa}. For details, we refer the reader to these two articles.

\begin{figure}
\centering
\vspace*{-1.7cm}
\hfill
\raisebox{1.7cm}{\text{$a)\ $}}\includegraphics[height=2.1in]{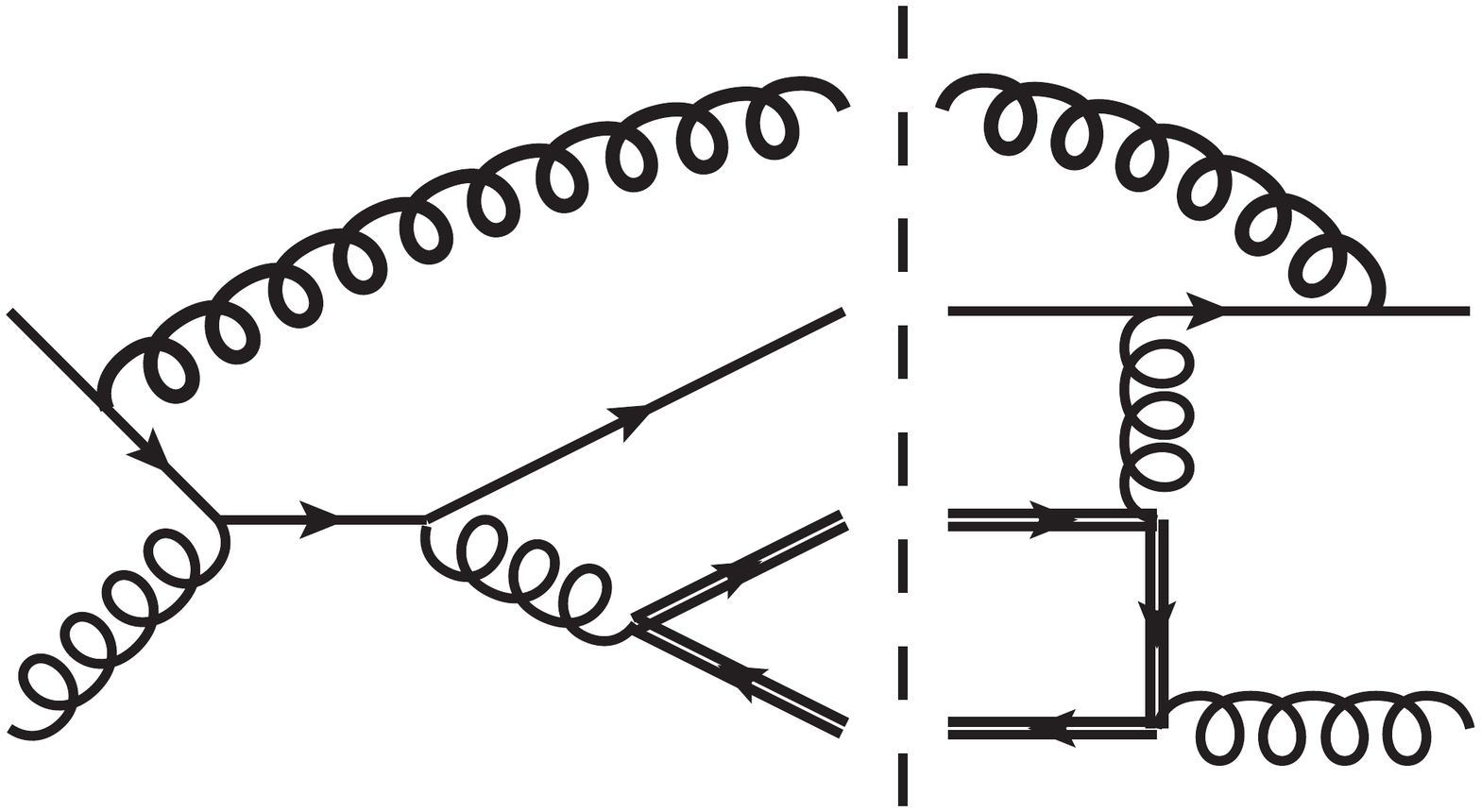}\hspace*{0.7cm}
\hfill
\raisebox{1.7cm}{\text{$b)\ $}}\raisebox{-0.1cm}{\includegraphics[height=2.1in]{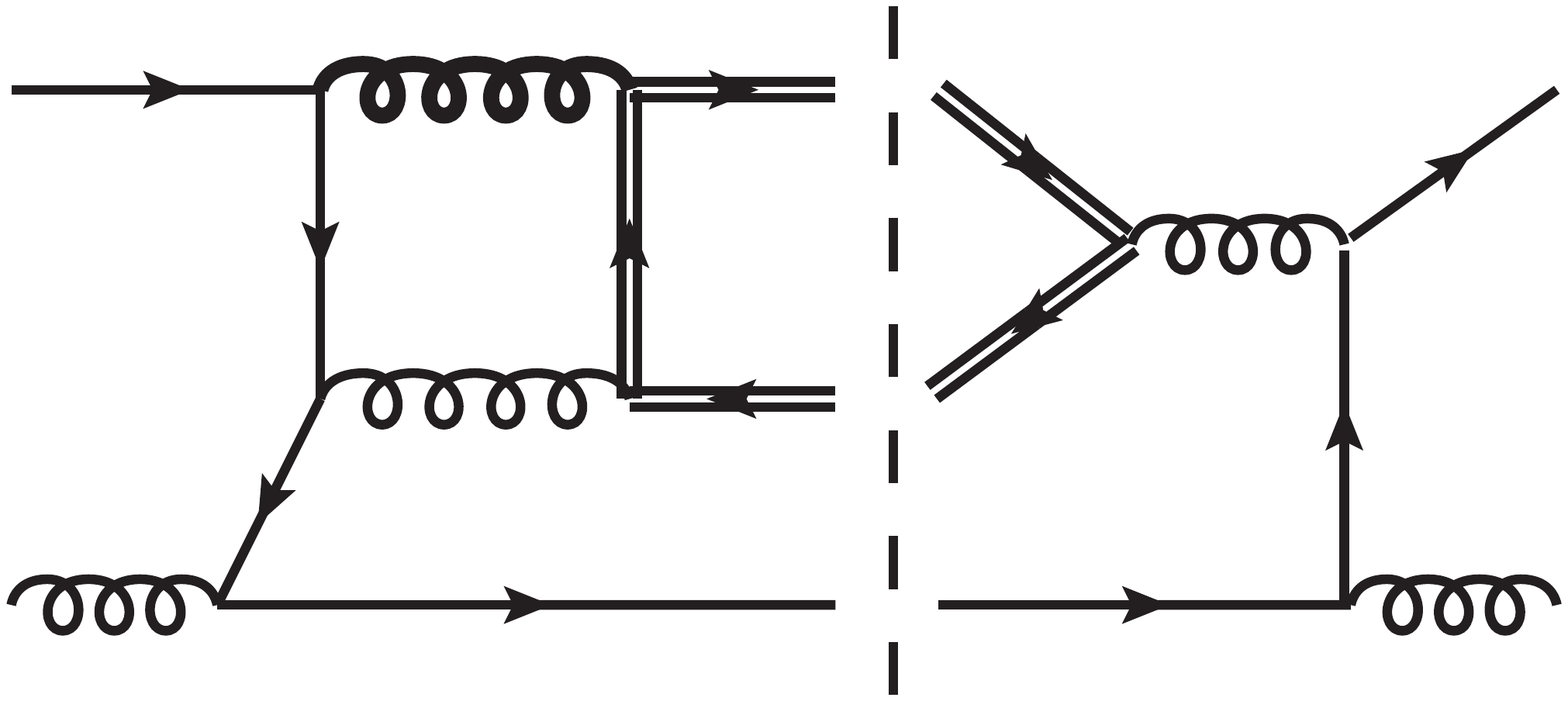}}\hspace*{0.5cm}
\hfill
\raisebox{1.7cm}{\text{$c)\ $}}\raisebox{-0.1cm}{\includegraphics[height=2.1in]{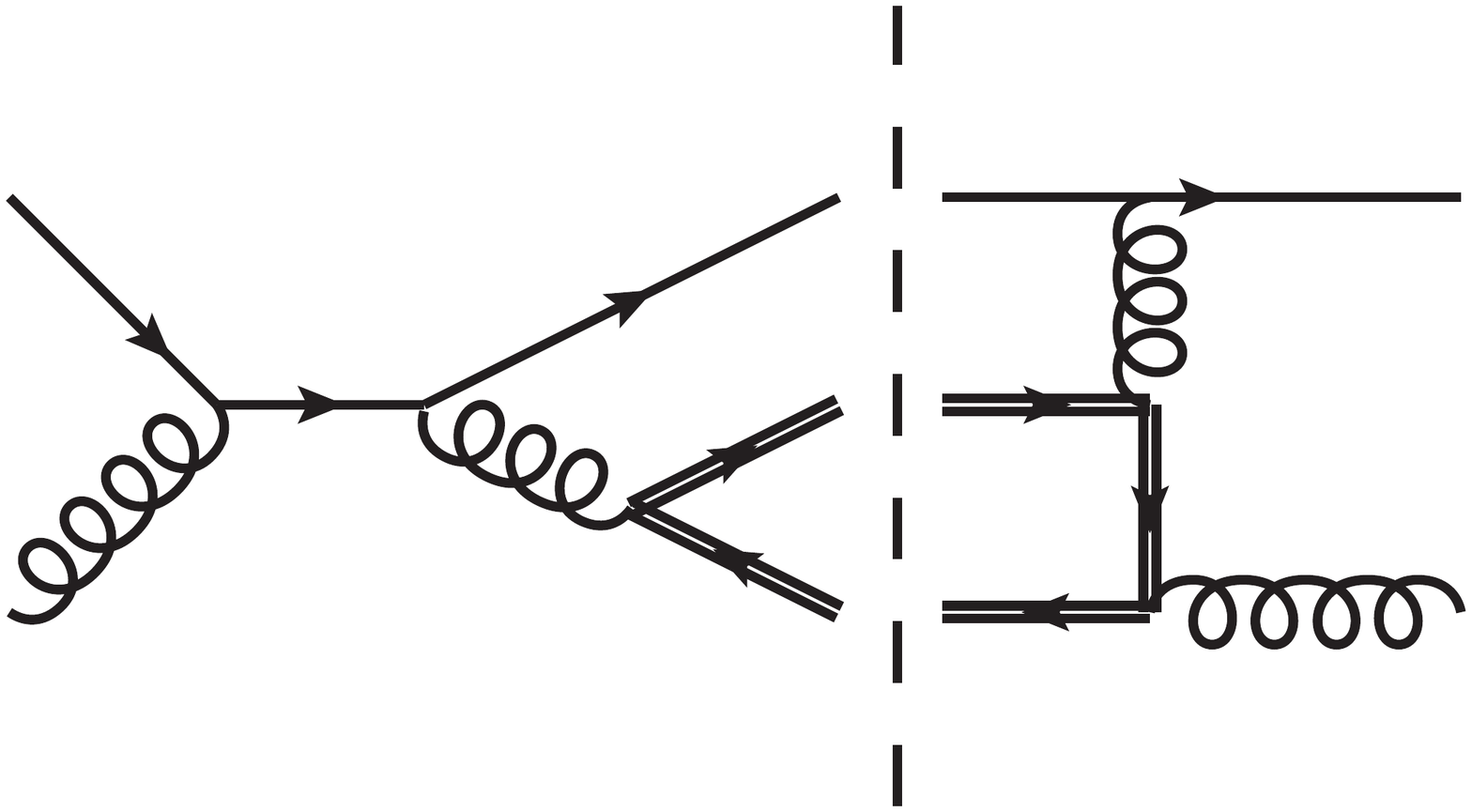}}
\hfill
\vspace{-1.5cm}
\caption{Feynman diagrams contributing to the charge asymmetry in the quark-gluon channel at NLO ($a$ and $b$) and at LO ($c$).}
\label{fig:diag1}
\end{figure}

The {\it energy asymmetry} in top-pair production requires the detection of an associated jet. In the process $pp\to t\bar t j$, the energy asymmetry is defined as~\cite{Berge:2013xsa}
\begin{equation}\label{eq:ea}
A_E(\theta_j)\equiv \frac{\sigma_{t\bar t j}(\theta_j,\Delta E > 0) - \sigma_{t\bar t j}(\theta_j,\Delta E < 0)}{\sigma_{t\bar t j}(\theta_j,\Delta E > 0) + \sigma_{t\bar t j}(\theta_j,\Delta E < 0)} \equiv\frac{\sigma_A(\theta_j)}{\sigma_S(\theta_j)}.
\end{equation}
Here $\Delta E = E_t-E_{\bar t}$ is the difference between the top and antitop energies. The angle $\theta_j$ is defined as the scattering angle of the hardest jet with respect to the direction of the incoming parton $p_1$ of the process $p_1 p_2\to t\bar t j$. Both $\Delta E$ and $\theta_j$ are defined in the $t\bar t j$ rest frame, which at LO corresponds with the parton CM frame. At the parton level, the energy asymmetry is generated mainly from quark-gluon initial states. An exemplary Feynman diagram contributing to the energy asymmetry at LO is shown in Fig.~\ref{fig:diag1}~$c$. At this level, the energy asymmetry is equivalent to a forward-backward asymmetry of the quark-jet in the top-antitop rest frame. Contributions to $A_E(\theta_j)$ from the quark-antiquark initial state are small and vanish when integrated over a range symmetric around $\theta_j = \pi/2$.\footnote{In our numerical analysis, we take quark-antiquark contributions into account.} The total energy asymmetry $A_E$ is obtained by integrating $\sigma_A(\theta_j)$ and $\sigma_S(\theta_j)$ in (\ref{eq:ea}) over the range $\theta_j \in [0,\pi]$.

The jet angular distribution $A_E(\theta_j)$ is shown in the upper right panel of Figure~\ref{fig:fo-lo-nlo}. As one can see, $A_E(\theta_j)$ is symmetric in $\theta_j \leftrightarrow \pi - \theta_j$, with a minimum at $\theta_j = \pi/2$. However, either of the two partonic contributions $qg\to t\bar t j$ and $gq\to t\bar t j$ has its minimum where the jet is emitted in the direction of the incoming quark, i.e., for $\theta_j < \pi/2$ and $\theta_j > \pi/2$, respectively. This feature is due to the kinematic distribution of the quark jet in the partonic process $qg\to t\bar t q$ at LO, which favors the quark direction (see also Reference~\cite{Berge:2013xsa}). This distribution translates to the hadron level, since the quark carries a larger fraction of the proton momentum along the beam axis than the gluon. We will ``guess'' the incoming quark's direction by exploiting the fact that the final state in quark-gluon interactions tends to be boosted in the quark direction, and thereby enhance the quark-gluon fraction of events against the gluon-gluon background. We quantify this boost in terms of the rapidity of the top-antitop-jet CM frame, $y_{t\bar t j}$, which at the LO is connected with the proton momentum fractions of $p_1$ and $p_2$ via $y_{t\bar t j}=\ln(x_1/x_2)/2$. This allows us to define an {\it optimized energy asymmetry} as~\cite{Alte:2014toa}
\begin{equation}\label{eq:ea-opt}
A_E^{\rm opt}(\theta_j) = \frac{\sigma_A(\theta_j,y_{t\bar tj}>0)+\sigma_A(\pi-\theta_j,y_{t\bar tj}<0)}{\sigma_S(\theta_j,y_{t\bar tj}>0)+\sigma_S(\pi-\theta_j,y_{t\bar tj}<0)} \equiv\frac{\sigma_A^{\rm opt}(\theta_j)}{\sigma_S^{\rm opt}(\theta_j)},
\end{equation}
where $\theta_j$ and $\sigma_{S,A}(\theta_j)=\sigma_{t\bar t j}(\theta_j,\Delta E > 0)\pm \sigma_{t\bar t j}(\theta_j,\Delta E < 0)$ are defined as in (\ref{eq:ea}), with an additional cut on the boost, $y_{t\bar t j}$. By associating forward ($y_{t\bar t j} > 0$) and backward ($y_{t\bar t j} < 0$) boosted final states with forward ($\theta_j \in [0,\pi/2]$) and backward ($\pi - \theta_j \in [\pi/2,\pi]$) scattered jets, we combine the maximum contributions from the $qg$ and $gq$ channels in the range $\theta_j \in [0,\pi/2]$. In the lower right panel of Figure~\ref{fig:fo-lo-nlo}, we show the jet angular distribution of the optimized energy asymmetry. The minimum of $A_E^{\text{opt}}(\theta_j)$ lies around $\theta_j\simeq 2\pi/5$ and is deeper than for $A_E(\theta_j)$.\footnote{Since $A_E^{\text{opt}}(\theta_j)$ is tailored to the kinematics of the partonic quark-gluon channel, quark-antiquark contributions deplete its minimum~\cite{Berge:2013xsa}. However, this effect is small due to the low quark-antiquark abundance in proton-proton collisions at the LHC.} In our numerical analysis in Section~\ref{sec:kinematics}, we will focus on this optimized energy asymmetry.

Occasionally, we will refer to the widely investigated {\it rapidity asymmetries} at the Tevatron,
\begin{equation}\label{eq:ra}
A_y^{(j)} = \frac{\sigma_{t\bar t(j)}(\Delta y > 0) - \sigma_{t\bar t(j)}(\Delta y < 0)}{\sigma_{t\bar t(j)}(\Delta y > 0) + \sigma_{t\bar t(j)}(\Delta y < 0)},\qquad \Delta y = y_t - y_{\bar t},
\end{equation}
and at the LHC,
\begin{equation}\label{eq:ra}
A_{|y|}^{(j)} = \frac{\sigma_{t\bar t(j)}(\Delta |y| > 0) - \sigma_{t\bar t(j)}(\Delta |y| < 0)}{\sigma_{t\bar t(j)}(\Delta |y| > 0) + \sigma_{t\bar t(j)}(\Delta |y| < 0)},\qquad \Delta |y| = |y_t| - |y_{\bar t}|.
\end{equation}
The top and antitop rapidities, $y_t$ and $y_{\bar t}$, are defined in the laboratory frame. The script $j$ is used if the top-antitop pair is produced in association with a hard jet. The rapidity asymmetry is generated from the quark-antiquark initial state and, unlike the energy asymmetry, does not require a hard jet in the final state. In the presence of an observed jet, however, $A_y^j$ is strongly reduced by QCD corrections at NLO~\cite{Dittmaier:2007wz,Dittmaier:2008uj,Melnikov:2010iu}, as we mentioned in the introduction. This fact raises the question if a similar effect occurs for the energy asymmetry. In what follows, we will discuss QCD corrections to the energy asymmetry in detail and explain why their impact is moderate.

\section{Energy asymmetry in QCD at NLO}\label{sec:nlo}
\begin{figure}[!tb]
\centering
\vspace*{-0.5cm}
\begin{tabular}{cc}
\includegraphics[scale=0.5]{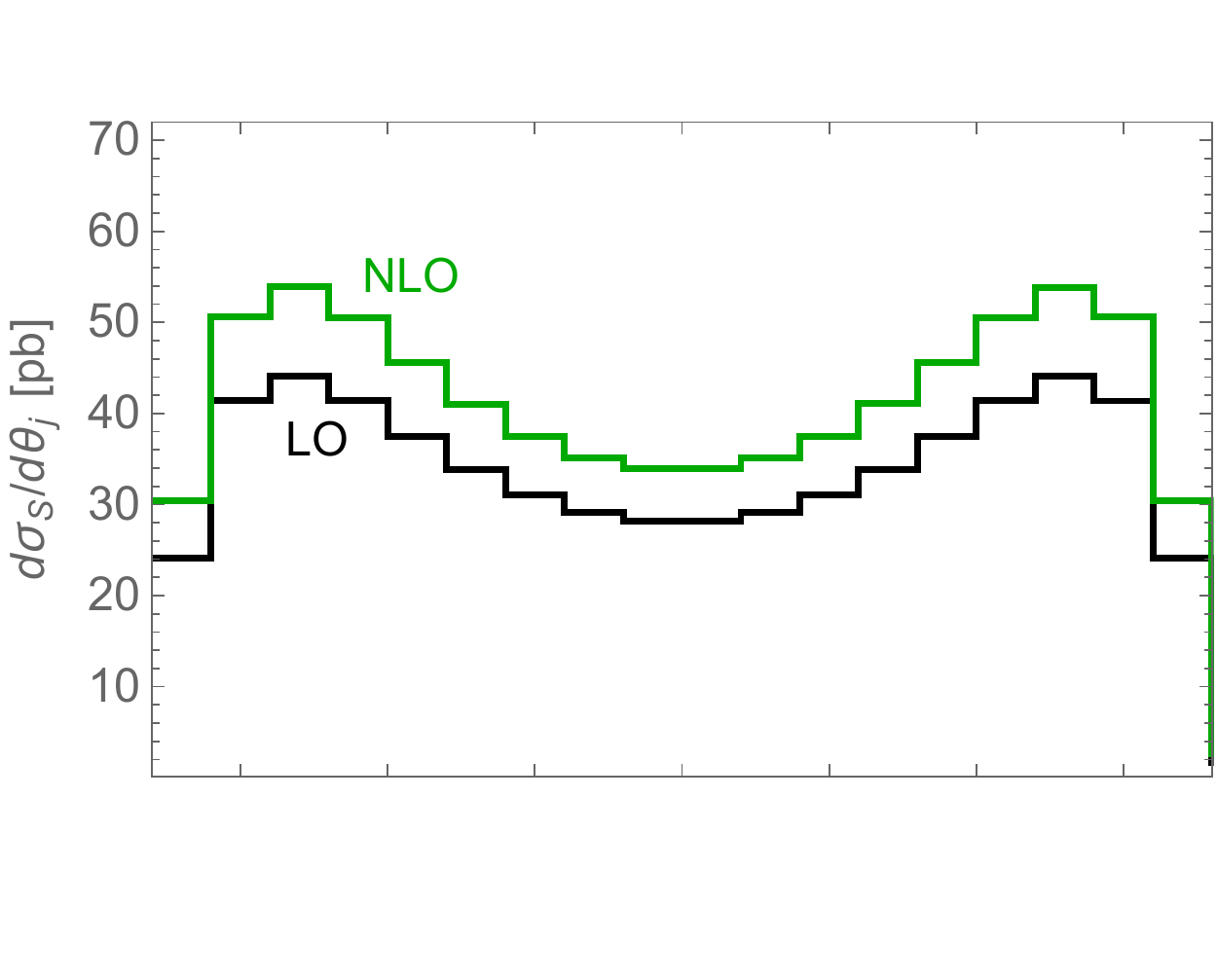}\hspace*{0.5cm} & \includegraphics[scale=0.5]{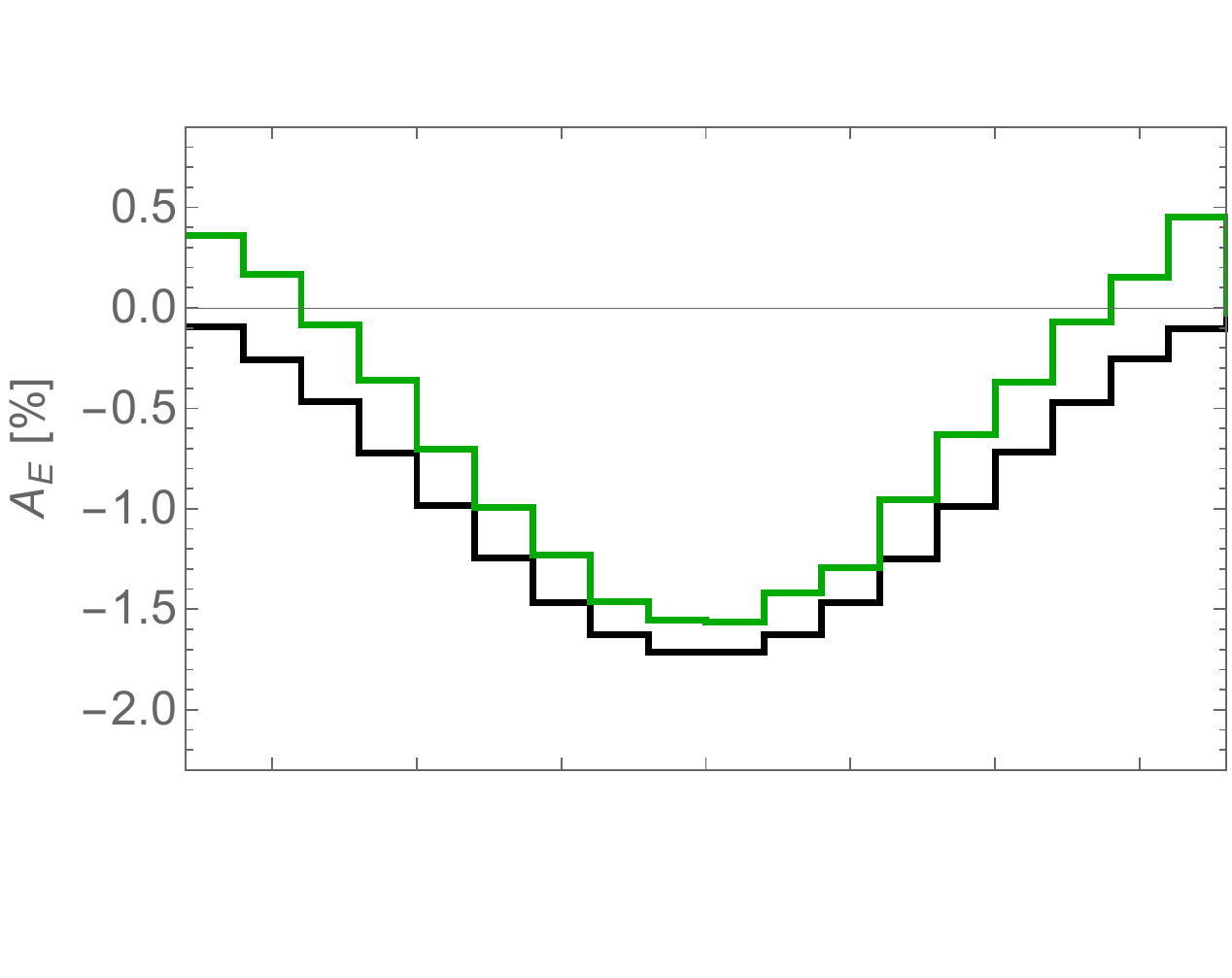}\\[-1.7cm]
\hspace*{0.14cm}\includegraphics[scale=0.5]{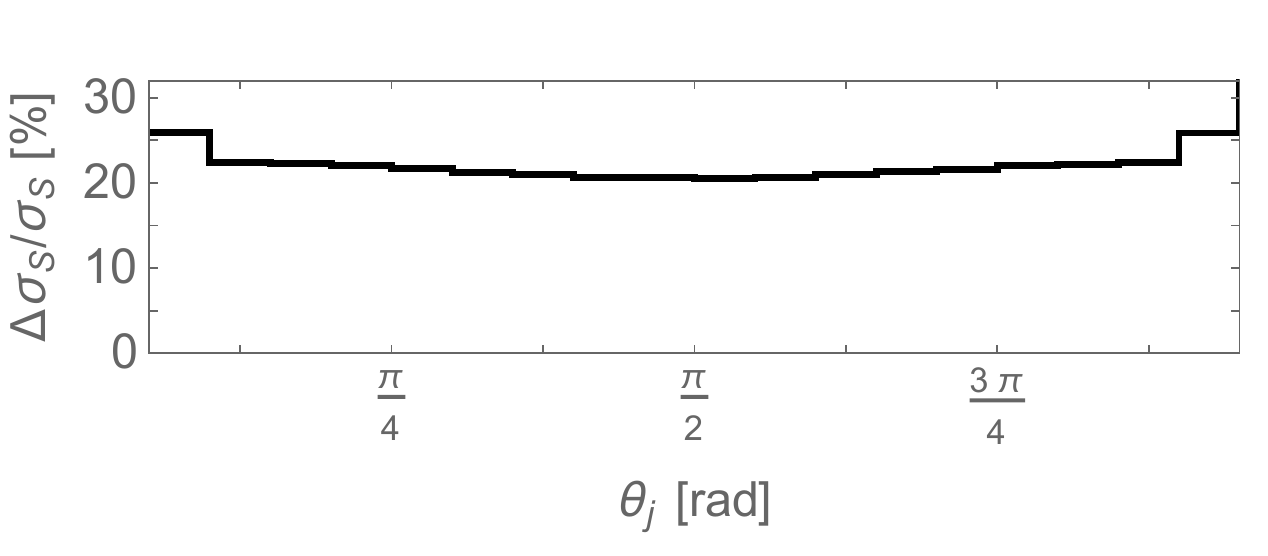}\hspace*{0.5cm} & \hspace*{0.32cm}\includegraphics[scale=0.5]{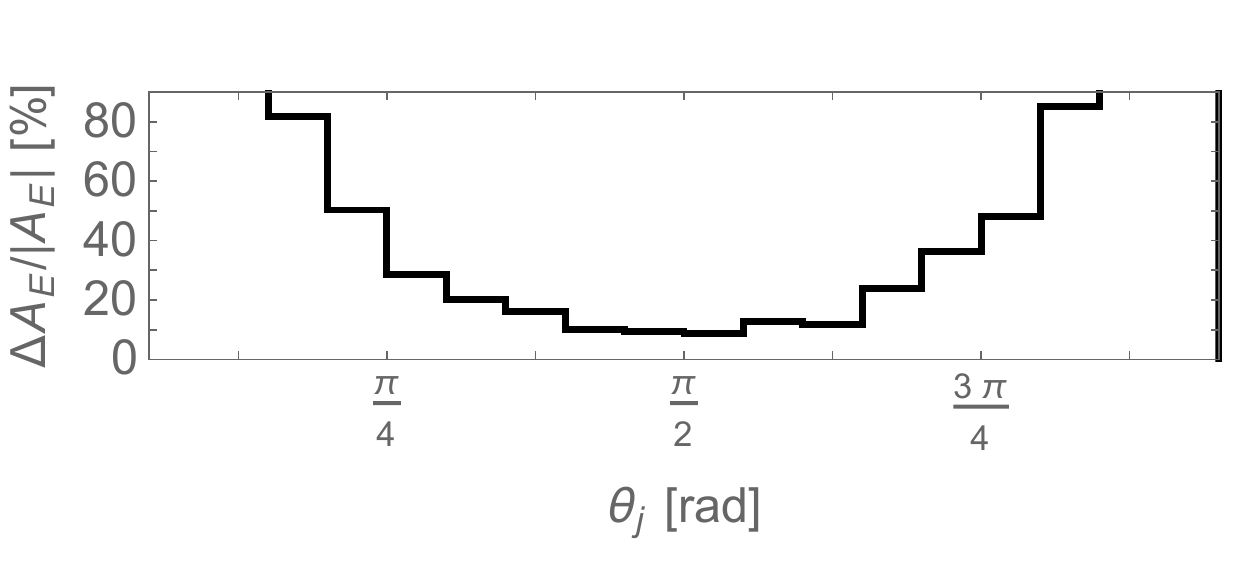}\\
\includegraphics[scale=0.5]{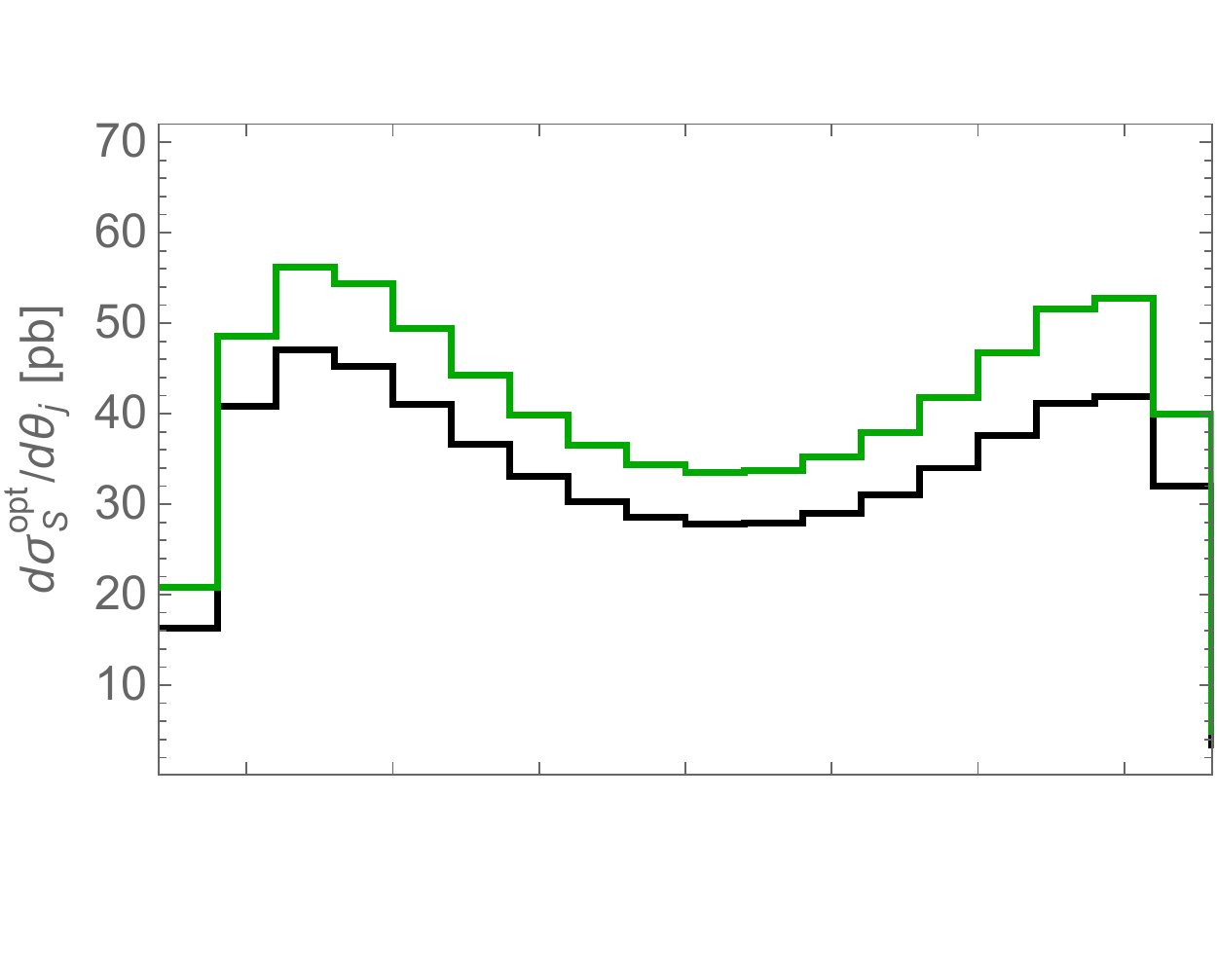}\hspace*{0.5cm} & \includegraphics[scale=0.5]{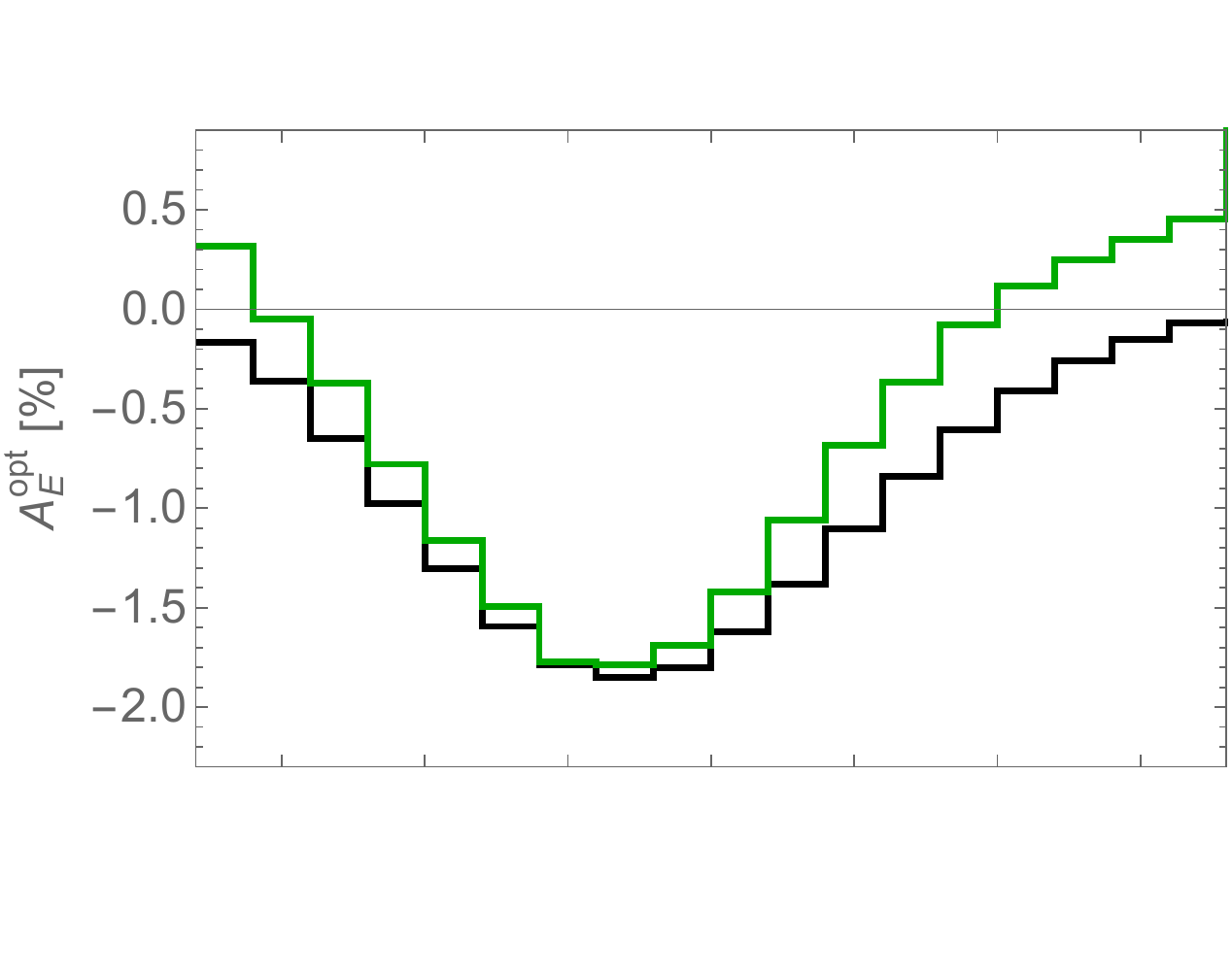}\\[-1.73cm]
\hspace*{0.12cm}\includegraphics[scale=0.5]{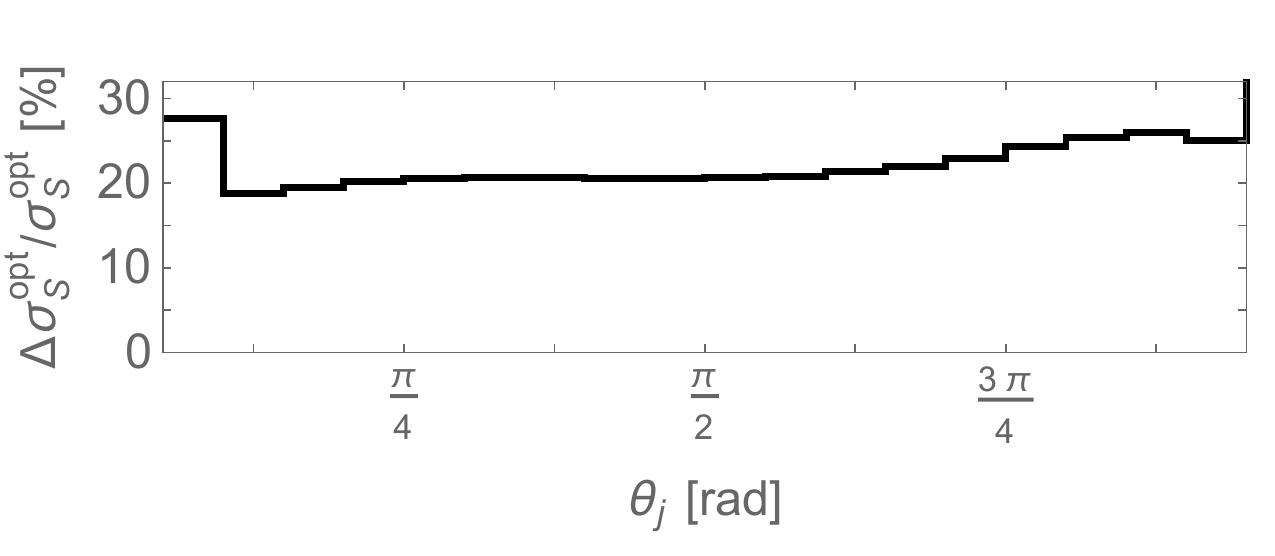}\hspace*{0.5cm} & \hspace*{0.32cm}\includegraphics[scale=0.5]{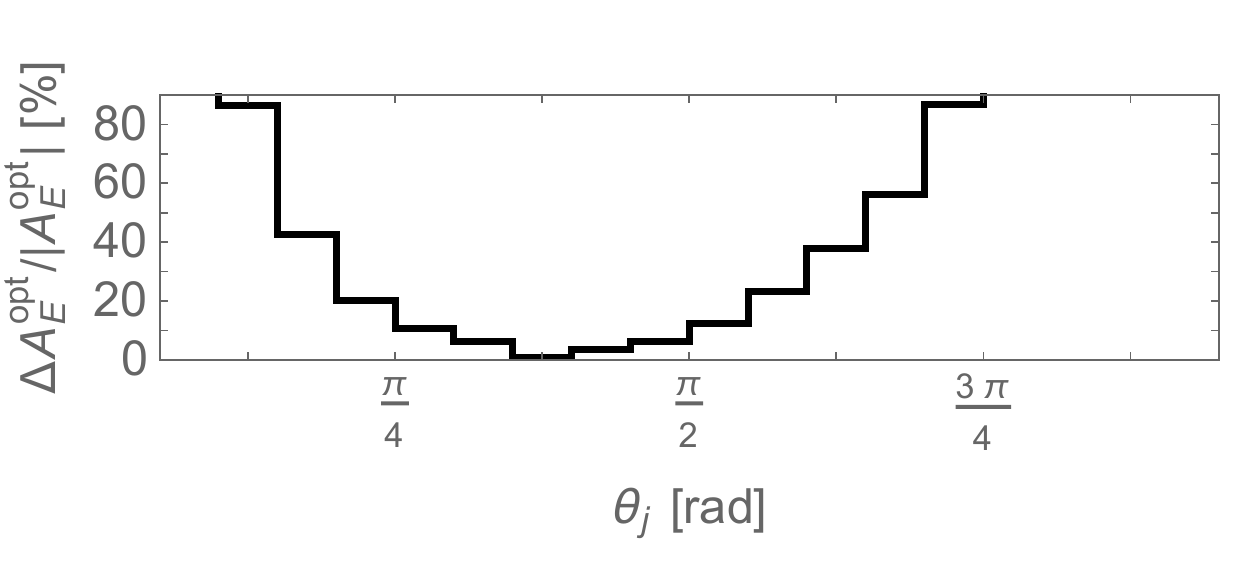}
\end{tabular}\vspace*{-0.1cm}
\caption{Cross section (left column) and charge asymmetry (right column) in $pp\to t\bar t j$ at $\sqrt{s}=13\,\text{TeV}$ as a function of the jet angle $\theta_j$. Fixed-order QCD results are shown at LO (black) and NLO (green) for the energy asymmetry (upper row) and the optimized energy asymmetry (lower row). Smaller panels display the relative effect of NLO corrections on the observable $O$, $\Delta O/O\equiv (O_{\rm NLO}-O_{\rm LO})/O_{\rm LO}$. Kinematic cuts of $p_T^j>100\,\text{GeV}$ and $|y_j| < 2.5$ have been applied. See Section~\ref{sec:lhc} for details.}\label{fig:fo-lo-nlo}
\end{figure}

The properties of the energy asymmetry at NLO are closely related to the kinematics of the observed jet. We can obtain a qualitative understanding by analyzing the cross section and the asymmetry in the limit of soft and collinear jet emission. The jet distribution of the cross section, $d\sigma_S/d\theta_j$, is displayed in the upper left panel of Figure~\ref{fig:fo-lo-nlo}. 
The maxima of the distribution are due to a double-logarithmic enhancement in the soft-collinear region, $\sigma_S\sim\log^2(m_t/p_T^j)$, which is regularized by a cut on the transverse momentum of the jet, $p_T^j$. In turn, the asymmetric cross section at LO, $\sigma_A^{\rm LO}$, is not logarithmically enhanced in this limit~\cite{Kuhn:1998kw}. At the LO, the energy asymmetry thus scales as
\begin{equation}\label{eq:ea-lo}
A_E^{\rm LO}=\frac{\sigma_A^{\rm LO}}{\sigma_S}\sim \log^{-2}\left(\frac{m_t}{p_T^j}\right).
\end{equation} 
At the NLO, $\sigma_A$ receives two types of contributions, represented by the diagrams in Fi\-gures~\ref{fig:diag1}\,$a$ and $b$. We assume that the observed jet stems from the quark in the final state. As we will see, this is true for the majority of jet-associated top-pair events. In Figure~\ref{fig:diag1}\,$a$, the asymmetry is generated by the same process as in the LO diagram of Figure~\ref{fig:diag1}\,$c$, with an additional gluon emission shown as a thick curly line. This type of NLO contributions to the asymmetry has the same asymptotic $p_T^j$ dependence as the LO contribution,
\begin{equation}\label{eq:ea-nlo-a}
A_E^{\rm NLO,a}=\frac{\sigma_A^{\rm NLO,a}}{\sigma_S}\sim \alpha_s\log^{-2}\left(\frac{m_t}{p_T^j}\right).
\end{equation}
In the second type of NLO contributions, the asymmetry is not generated by the quark-jet, but by the exchange of a virtual gluon, as shown in Figure~\ref{fig:diag1}\,$b$. This contribution is logarithmically enhanced in the limit of collinear jet emission, $\sigma_A^{\rm NLO,b}\sim \alpha_s\log(m_t/p_T^j)$. The normalized asymmetry therefore depends on the jet momentum as
\begin{equation}\label{eq:ea-nlo-b}
A_E^{\rm NLO,b} = \frac{\sigma_A^{\rm NLO,b}}{\sigma_S}\sim \alpha_s\log^{-1}\left(\frac{m_t}{p_T^j}\right).
\end{equation}
The collinear enhancement $\sigma_A^{\rm NLO,b}\sim  \alpha_s\log(m_t/p_T^j)$ is common to all contributions that can be traced back to an asymmetry in the inclusive process $q\bar q\to t\bar t$, where the parton splitting $g\to q\bar q$ has been absorbed in the parton distribution of the incoming quark.

In summary, the total energy asymmetry at NLO is given by
\begin{equation}\label{eq:ea-nlo-tot}
A_E^{\rm NLO} = \frac{\sigma_A^{\rm LO}+\sigma_A^{\rm NLO,a}+\sigma_A^{\rm NLO,b}}{\sigma_S}\sim \log^{-2}\left(\frac{m_t}{p_T^j}\right)\times \Big[1+\mathcal{O}(\alpha_s)+\mathcal{O}(\alpha_s\log\left(\frac{m_t}{p_T^j}\right))\Big].
\end{equation}
In the limit $p_T^j\to 0$, the energy asymmetry vanishes as expected for inclusive top-antitop production. For small $p_T^j$, NLO corrections of type $b$ can be important. In our analysis, we will apply a lower cut on the transverse momentum of the hardest jet, $p_T^j > 100\,\text{GeV}$. Therefore, in the kinematic region of our interest the asymmetry is not logarithmically suppressed and NLO QCD corrections are moderate.\footnote{The rapidity asymmetry in $pp\to t\bar t j$ has a different $p_T^j$ dependence. At small $p_T^j$, the LO contribution to $A_y^j$ is logarithmically suppressed as $\log^{-1}(m_t/p_T^j)$, whereas NLO contributions from hard degrees of freedom do not logarithmically depend on $p_T^j$~\cite{Melnikov:2010iu}. For moderate cuts on $p_T^j$, the rapidity asymmetry thus receives large (positive) NLO corrections, which partially cancel the (negative) asymmetry at LO. A harder cut on $p_T^j$ helps reducing these NLO corrections to $A_y^j$~\cite{Alioli:2011as}.} Similar considerations apply for the optimized energy asymmetry, $A_E^{\rm opt}$.

\section{Numerical analysis for the LHC at 13 TeV}\label{sec:lhc}
With these qualitative considerations in mind, we turn to a quantitative analysis of the energy asymmetry. Our results have been obtained numerically with Monte Carlo methods. We have used the program {\tt MadGraph5$\_$aMC@NLO} version 2.4.0~\cite{Alwall:2014hca}, which allows us to compute hadronic cross sections and differential distributions of the process $pp\to t\bar t j$ up to NLO in a largely automized way. Hard matrix elements have been folded with parton distribution functions (PDFs) using the interpolator {\tt LHAPDF} 6.1.6~\cite{Buckley:2014ana}. We have worked in a factorization scheme with five active quark flavors, where all quarks but the top-quark are considered massless. Jets in the final state are found and ordered with respect to their transverse momentum using the program {\tt FastJet} release 3.1.3~\cite{Cacciari:2011ma} inside {\tt aMC@NLO}. Hadronization has not been included; in particular, top-quarks have been kept stable in the final state. If not specified otherwise, throughout our analysis we have used the setting in Table~\ref{tab:inputs}. We have performed 70 independent runs, requesting a precision of $5\permil$ for the cross section resulting from each run. The so-obtained events have been analyzed using our private code.
\begin{table}[!tb]
\centering
\renewcommand{\arraystretch}{1.5}
\begin{tabular}{|l|}
\hline
$\alpha_s(m_Z)=0.118$; $m_t=173.2\gev$; $\sqrt{s}=13\,\text{TeV}$\\
default parton distributions: {\tt NNPDF 3.0} NLO\\
default scale choice: $\mu_R=\mu_F=m_t$\\
anti-kt jet clustering algorithm; jet radius $\Delta R=0.5$\\
cuts on hardest jet: $p_T^j > 100\gev$, $|y_j| < 2.5$\\
\hline
\end{tabular}
\caption{Inputs for the numerical analysis.}
\label{tab:inputs}
\end{table}

In Figure~\ref{fig:fo-lo-nlo}, we show the cross sections $\sigma_S$, $\sigma_S^{\rm opt}$ and energy asymmetries $A_E$, $A_E^{\rm opt}$ defined in Section~\ref{sec:observables}.  We present all observables as functions of the jet scattering angle in the $t\bar t j$ CM frame, $\theta_j$. Jet distributions at LO and NLO are displayed in black and green, respectively. The smaller panels show the relative impact of NLO QCD contributions on the observable $O$, i.e., $\Delta O/O \equiv (O_{\rm NLO}-O_{\rm LO})/O_{\rm LO}$. Kinematic cuts $p_T^j > 100\,\text{GeV}$ and $|y_j| < 2.5$ on the transverse momentum and the rapidity of the jet with the highest transverse momentum have been applied here and (unless said otherwise) throughout our analysis.\footnote{The rapidity $y_j$ is defined in the laboratory frame.} These cuts regularize the observables in the soft and collinear regions. The strong cut on $p_T^j$ enhances the fraction of quark-gluon initial states in the total cross section to roughly $40\%$, which is important for suppressing the gluon-gluon background to the asymmetry. Furthermore, the cuts ensure that the hardest observed jet is most likely the quark-jet, rather than one of the typically softer jets from higher-order QCD effects to the production process. Another source of QCD radiation stems from top decays. However, such radiation is also typically softer than $100\,\text{GeV}$. We thus expect that the strong transverse momentum cut on the hardest jet is efficient in suppressing potential jet mis-identification through additional QCD radiation and necessary to prevent the energy asymmetry from being washed out by such effects.

At the LO, both asymmetries $A_E$ and $A_E^{\rm opt}$ are negative over the entire range of jet angles and vanish for $\theta_j = 0,\,\pi$.\footnote{Due to the smallness of the asymmetries near the endpoints, the results for $A_E(\theta_j)$ and $A_E^{\rm opt}(\theta_j)$ in the marginal bins are subject to statistical uncertainties. We therefore refrain from showing those bins here and in the following figures.} The energy asymmetry $A_E$ in the upper right panel of Figure~\ref{fig:fo-lo-nlo} has its minimum at $\theta_j = \pi/2$, where the hard jet is emitted perpendicular to the beam axis in the $t\bar t j$ CM frame~\cite{Berge:2012rc,Berge:2013xsa}. NLO contributions are positive; they reduce the magnitude of the asymmetry by a few relative percent at $\theta_j=\pi/2$ and grow towards the endpoints of the spectrum, where they ultimately drive the asymmetry to positive values. This behavior can be understood by looking at the two types of NLO contributions to the asymmetry discussed in Section~\ref{sec:nlo}. NLO contributions of type $a$ are expected to be negative and moderate, since they are induced by the same process that generates the asymmetry at LO. Around $\theta_j = 0,\pi$, these corrections are over-compensated by contributions of type $b$, which are positive and grow logarithmically for small jet angles. Type-$b$ contributions are of the same origin as virtual contributions to the asymmetry in inclusive top-antitop production. Indeed, a similar interplay between positive virtual and negative real contributions has previously been observed for the forward-backward asymmetry in inclusive top-antitop production~\cite{Kuhn:1998kw} and in jet-associated top-antitop production~\cite{Dittmaier:2007wz,Dittmaier:2008uj,Melnikov:2010iu}.

The optimized energy asymmetry $A_E^{\rm opt}$ has a somewhat different jet distribution, due to the cut on the boost $y_{t\bar t j}$, which tends to point in the direction of the incoming quark in quark-gluon interactions. The selection $y_{t\bar t j} > 0$ enhances the fraction of quark-gluon contributions to the cross section for $\theta_j \lesssim \pi/2$ and reduces it for $\theta_j \gtrsim \pi/2$ (and vice versa for gluon-quark contributions). Therefore, the distribution $d\sigma_S^{\rm opt}/d\theta_j$ is lopsided, as can be seen in the lower left panel of Figure~\ref{fig:fo-lo-nlo}. Due to the quark-gluon enhancement, $A_E^{\rm opt}$ (in the lower right panel) develops faster for small $\theta_j$ than $A_E$ and reaches its minimum around $\theta_j\simeq 2\pi/5$. The cut on $y_{t\bar t j}$ also has an impact on NLO effects. For contributions of type $a$ and $b$, the jet is preferentially emitted in and against the direction of the incoming quark, respectively. The (negative) type-$a$ NLO contributions to the asymmetry are thus enhanced for $\theta_j < \pi/2$, while (positive) type-$b$ contributions are enhanced for $\theta_j > \pi/2$. This explains why NLO contributions to $A_E^{\rm opt}$ grow faster at large jet angles than at small angles. Around the minimum, however, NLO corrections to $A_E^{\rm opt}$ are even smaller than for $A_E$. Beyond the NLO, no new classes of diagrams enter the process, meaning that NNLO contributions to the energy asymmetry are of either type $a$ and $b$. We therefore expect NNLO corrections to the asymmetry to be moderate.

In summary, the optimized energy asymmetry $A_E^{\rm opt}$ has two advantages over the energy asymmetry $A_E$. Firstly, its minimum is more pronounced due to the enhanced quark-gluon fraction in this slice of phase-space. And secondly, NLO corrections around the minimum are smaller, since type-$b$ effects are less important in this region of phase space. These observations lead us to focus on the optimized energy asymmetry in what follows. In Section~\ref{sec:kinematics}, we will show how to exploit these features to maximize the asymmetry by suitable cuts on the jet angle. Integrated over a region that is symmetric around $\theta = \pi/2$, $A_E$ and $A_E^{\rm opt}$ coincide per definition. With a smaller data set that does not allow for kinematic cuts, $A_E$ and $A_E^{\rm opt}$ are thus equally well motivated observables.

\begin{figure}[!t]
\centering
\vspace*{-0.5cm}
\begin{tabular}{cc}
\includegraphics[scale=0.5]{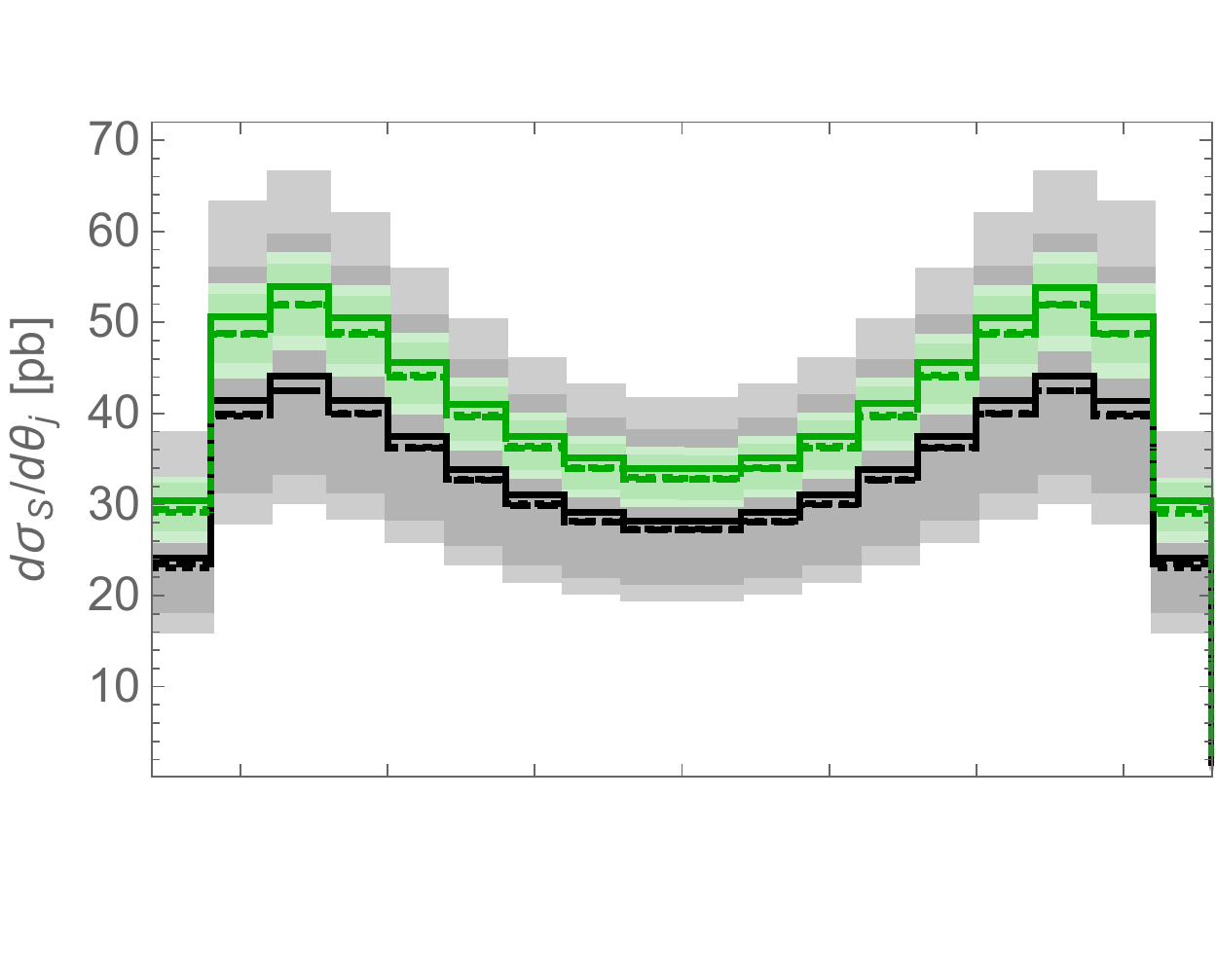}\hspace*{0.5cm} & \includegraphics[scale=0.5]{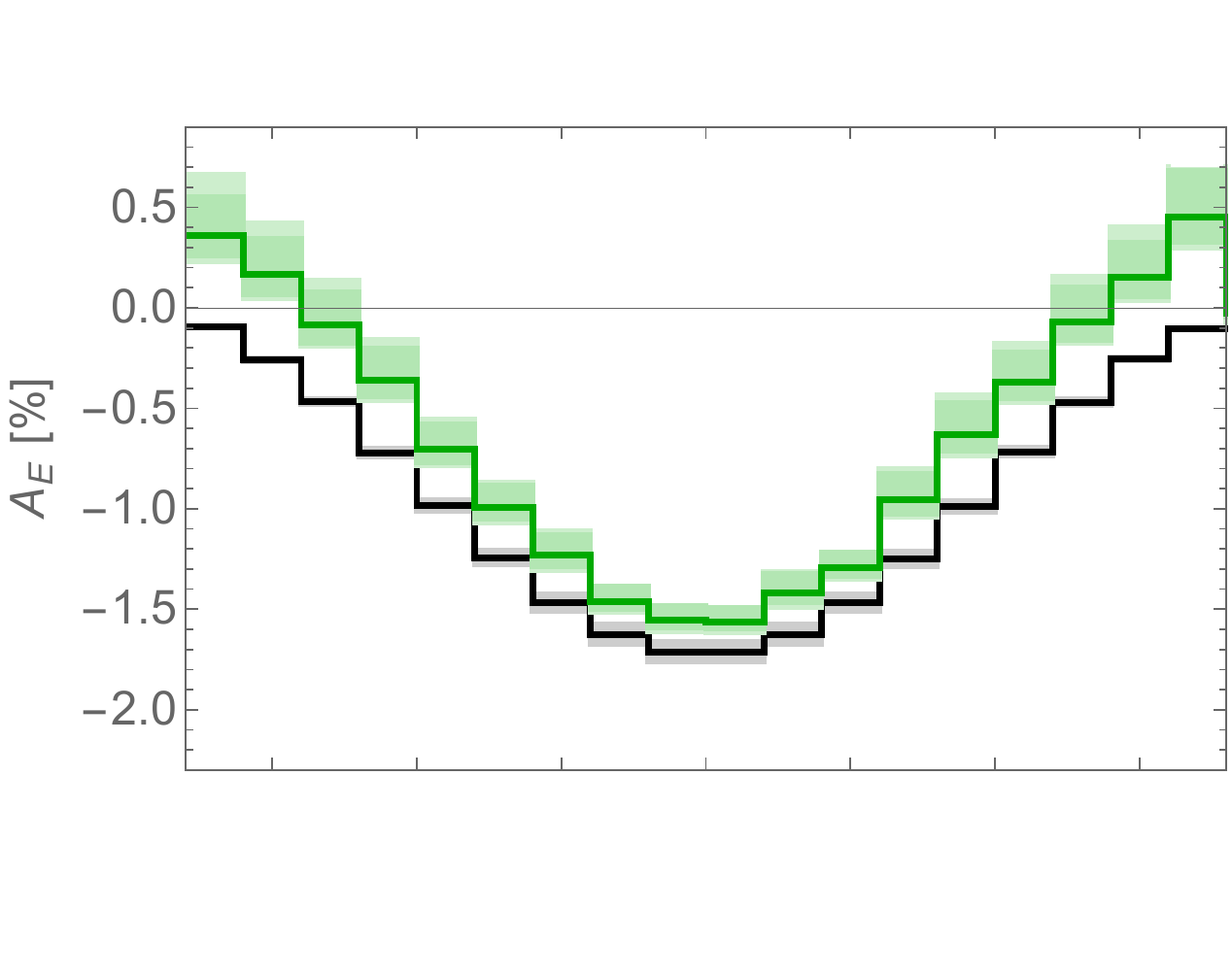}\\[-1.71cm]
\raisebox{-0.061cm}{\includegraphics[scale=0.522]{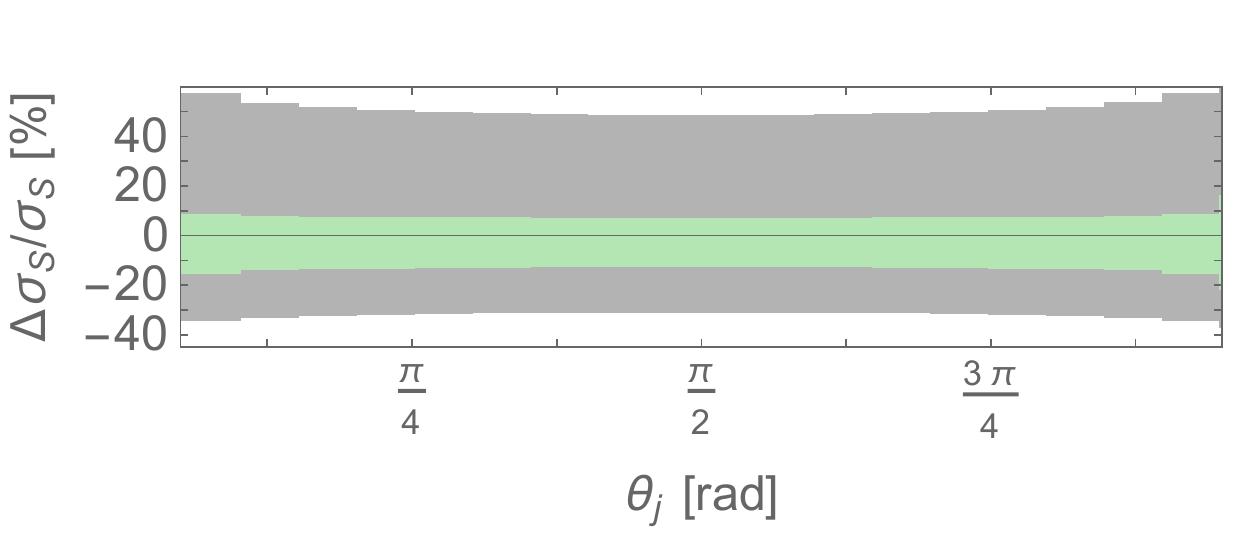}}\hspace*{0.53cm} & \hspace*{0.17cm}\includegraphics[scale=0.5]{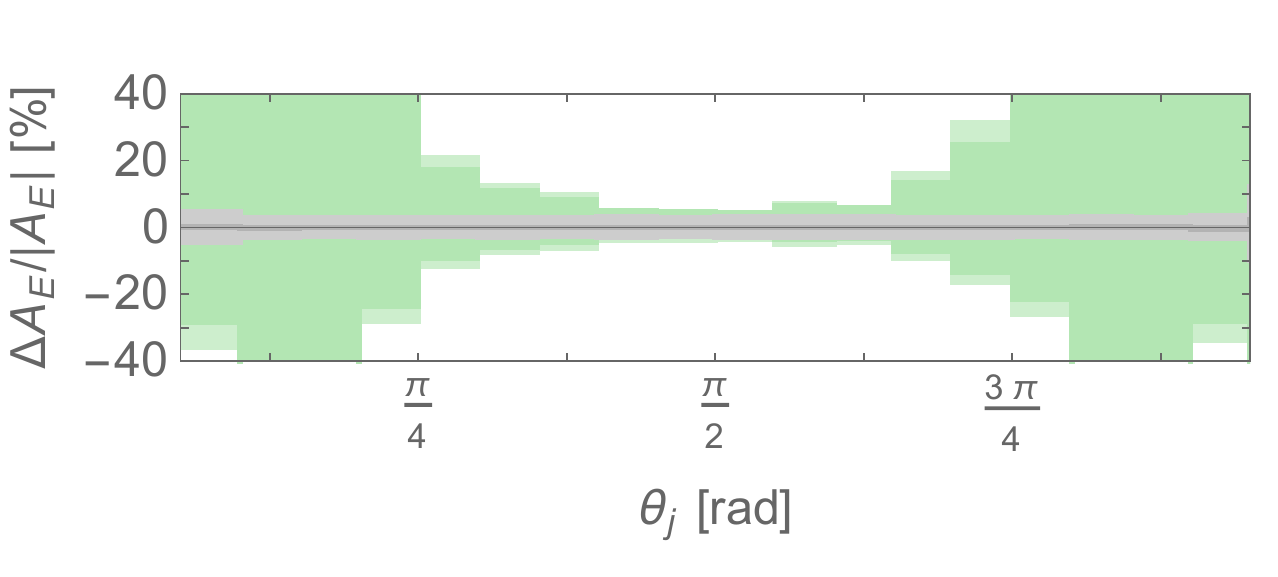}\\
\includegraphics[scale=0.5]{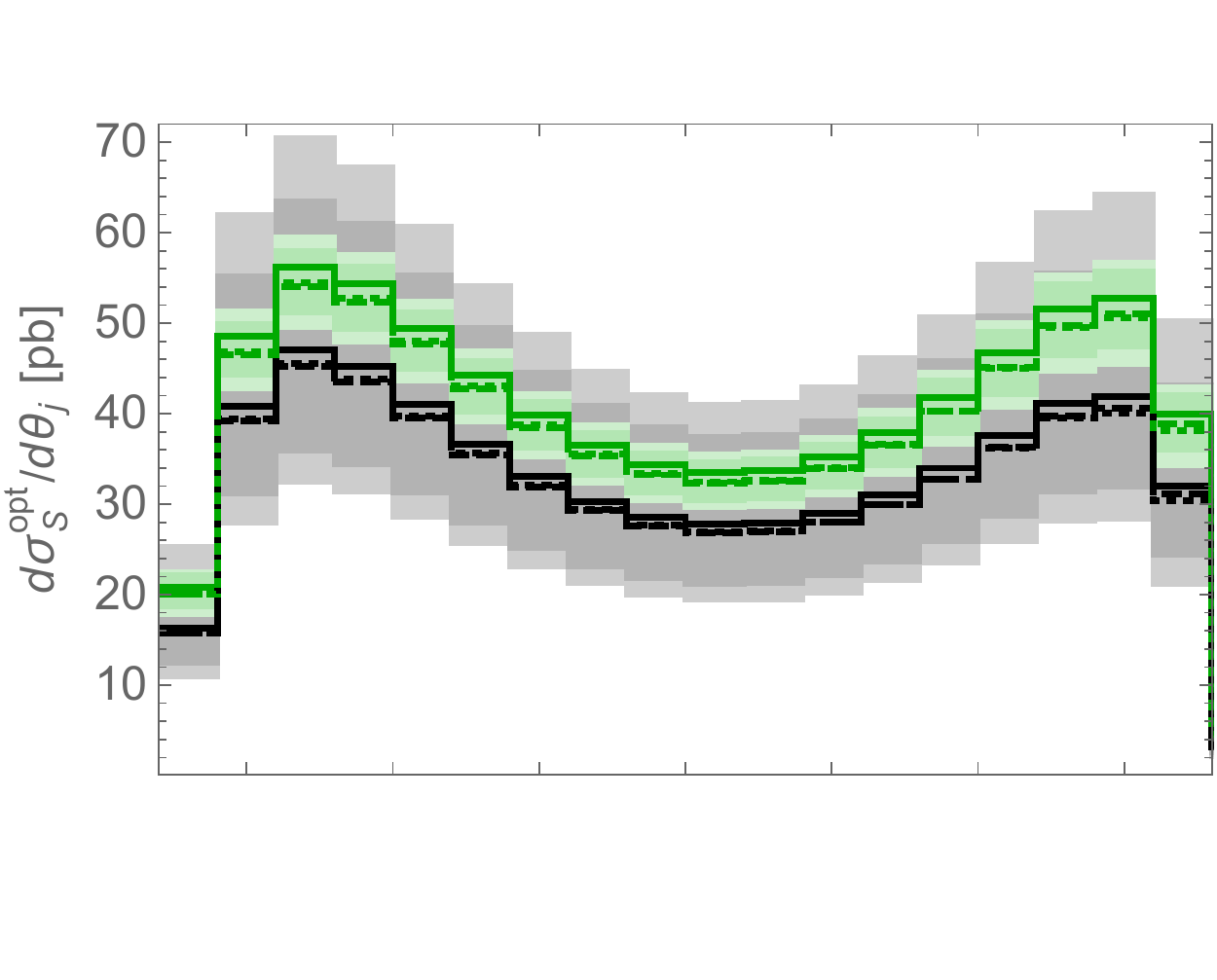}\hspace*{0.5cm} & \includegraphics[scale=0.5]{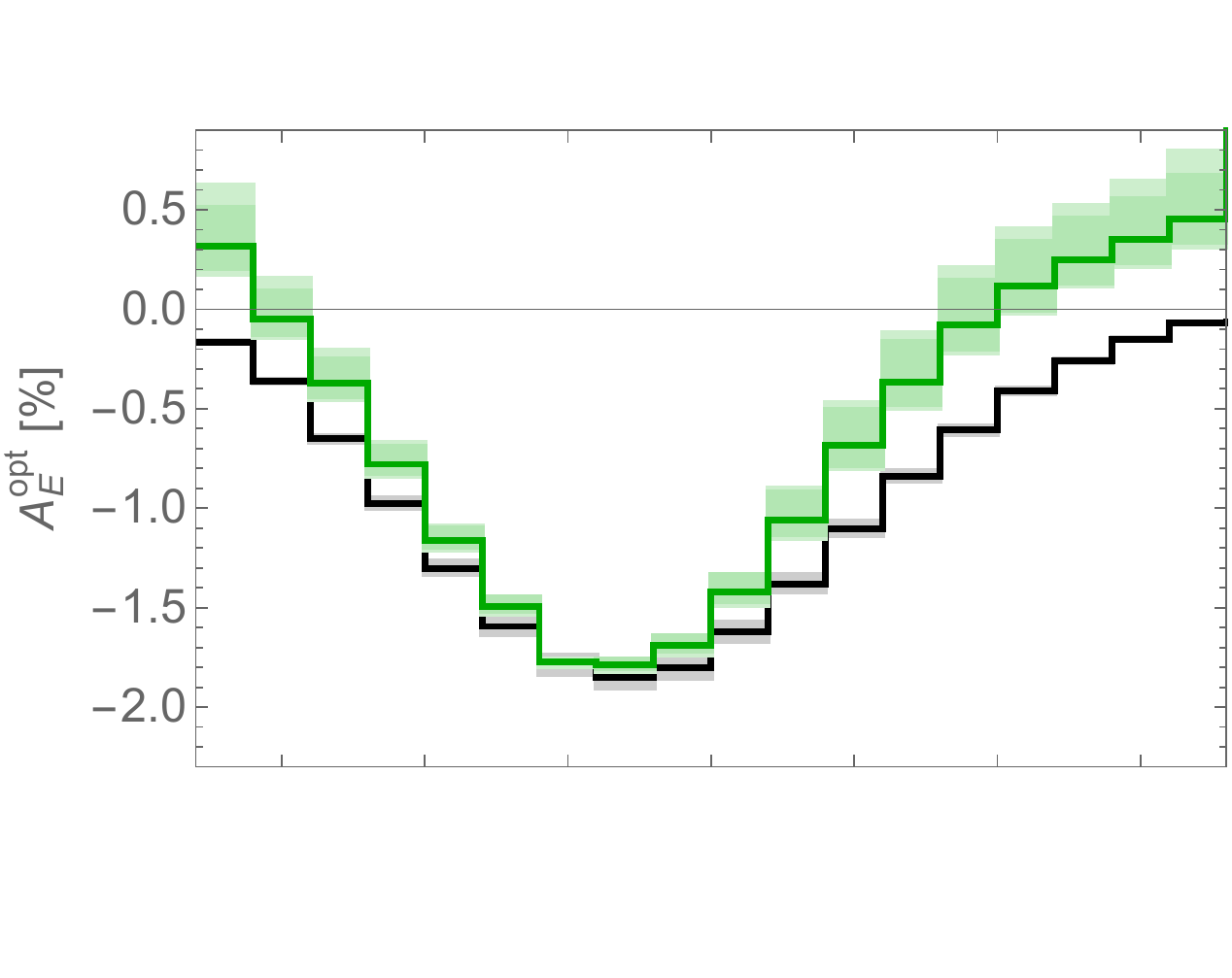}\\[-1.77cm]
\hspace{-0.02cm}\raisebox{0.01cm}{\includegraphics[scale=0.514]{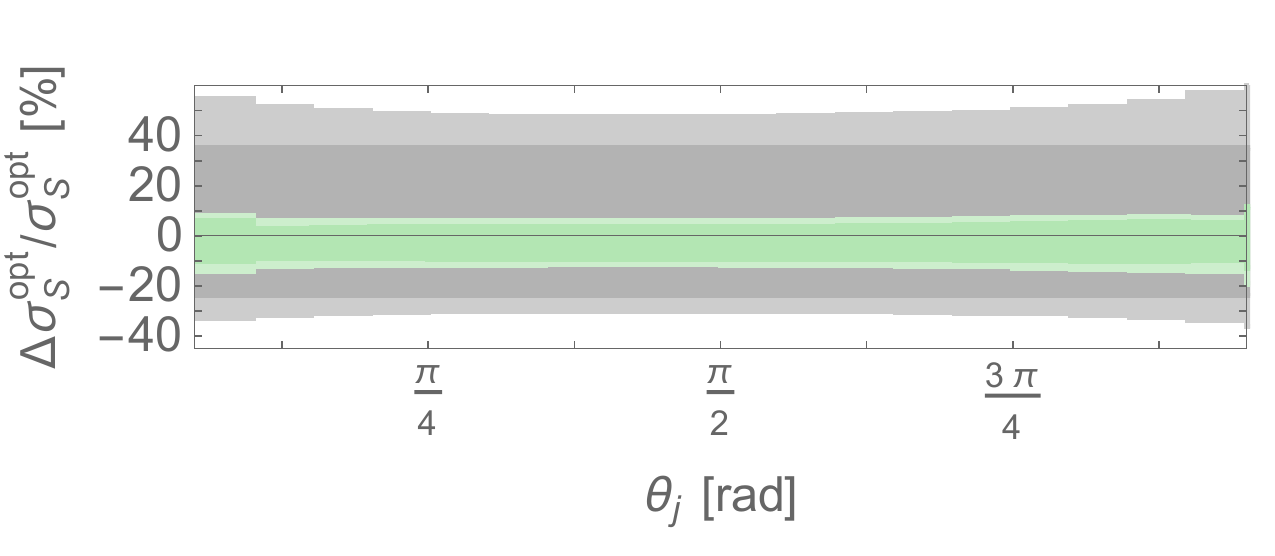}}\hspace*{0.54cm} & \hspace*{0.15cm}\includegraphics[scale=0.515]{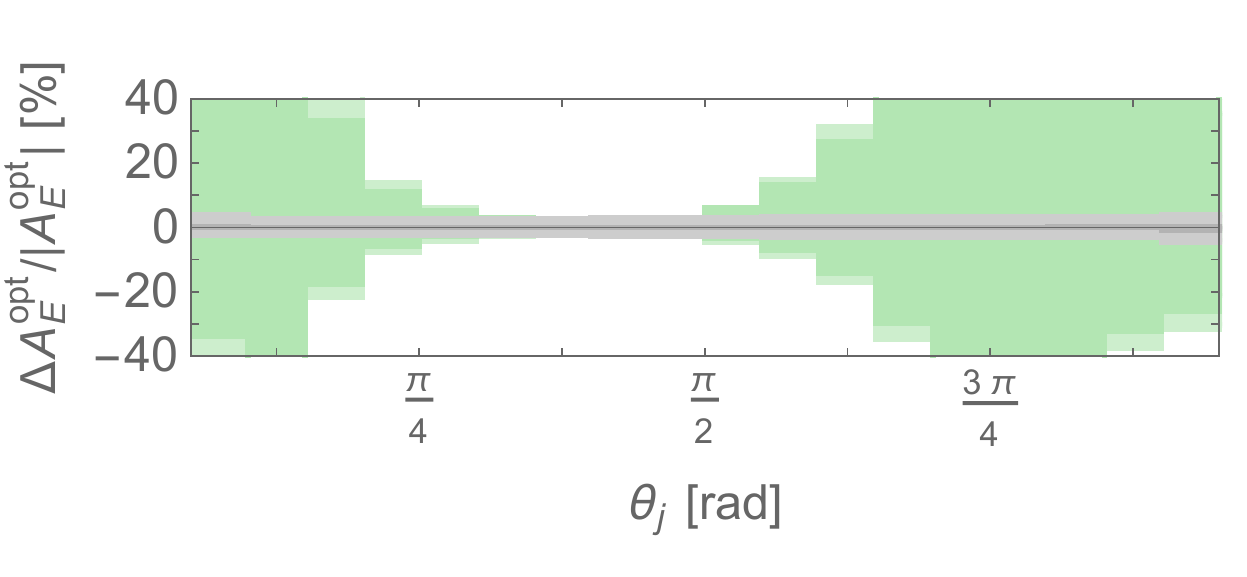}\\
\end{tabular}\vspace*{-0.1cm}
\caption{Scale and PDF variations of the cross section (left) and energy asymmetry (right) at LO (black) and NLO (green).  Central values for {\tt NNPDF}, {\tt CT14} and {\tt MMHT} PDFs are shown as plain, dashed, and dotted lines, respectively. Light (dark) bands show scale variations in the range $\mu_R,\mu_F\in [0.5,2]\,m_t$ ($\mu_R \in [0.5,2]\,m_t,\ \mu_F=m_t$) for the {\tt NNPDF} set; corresponding relative scale variations with respect to the central value are given in the smaller panels.}
\label{fig:scale}
\end{figure}

A few remarks on the precision of our predictions are in order. The dominant source of uncertainties stems from missing higher-order QCD corrections, which manifest themselves in a remnant dependence on the factorization and renormalization scales, $\mu_F$ and $\mu_R$. In Figure~\ref{fig:scale}, we display the cross sections and asymmetries from Figure~\ref{fig:fo-lo-nlo} with scale variations. Light-shaded bands show the envelope obtained when varying $\mu_R$ and $\mu_F$ independently in the range $[0.5,2]\,m_t$ in steps of $0.5\,m_t$. Dark-shaded bands show the variation of the renormalization scale only, $0.5\,m_t < \mu_{R} < 2\,m_t$, with $\mu_F = m_t$ kept fixed. The relative change of the observables with respect to the central values with $\mu_R=\mu_F=m_t$ (the plain line) is shown in the smaller panels.

At LO, the renormalization scale dependence enters the matrix elements through the running QCD coupling, resulting in $\sigma_{S,A}\sim \alpha_s^3(\mu_R)$. For the cross sections $\sigma_S$ and $\sigma_S^{\rm opt}$, we observe a strong reduction of scale uncertainties from about $40\%$ at LO to roughly $10\%$ when including NLO corrections. In the asymmetries $A_E$ and $A_E^{\rm opt}$, the renormalization scale dependence cancels completely between the numerator and denominator at LO. A small factorization scale dependence of about $5\%$ is left. At NLO, however, a renormalization scale dependence is re-introduced mainly through virtual type-$b$ contributions to the asymmetry, which does not cancel with the normalization. Since these virtual contributions grow lo\-ga\-rith\-mi\-cal\-ly in the collinear limit, scale uncertainties are larger at the endpoints of the jet distribution. However, around the minima of the asymmetries, scale uncertainties mostly cancel between $\sigma_A$ and $\sigma_S$, up to an uncertainty of a few percent only. This fact allows us to make a precise prediction of the energy asymmetry at NLO in the kinematic region near the minimum. Due to the shift by type-$b$ contributions at NLO and the partial cancellation of scale variations in the normalized asymmetry, the different scale uncertainties of $A_E$ and $A_E^{\rm opt}$ for LO and NLO predictions should not be interpreted in terms of perturbative convergence. Looking at the NLO asymmetry prediction by itself, however, we expect the scale uncertainties to give an estimate of missing higher-order QCD corrections, which will be of either type $a$ or type $b$.

A second source of uncertainties is due to PDFs. In Figure~\ref{fig:scale}, we show predictions of the jet distributions for three PDF sets, {\tt NNPDF 3.0}~\cite{Ball:2014uwa} (plain), {\tt CT14}~\cite{Dulat:2015mca} (dashed), and {\tt MMHT 2014}~\cite{Harland-Lang:2014zoa} (dotted), with $\alpha_s(m_Z)=0.118$ and $\mu_F=m_t=\mu_R$. We have used NLO PDFs throughout our analysis, i.e., as well for LO predictions. This allows us to better compare hard QCD effects at LO and NLO. From the figures, it is apparent that {\tt CT14} and {\tt MMHT} PDFs give almost identical predictions of the cross sections, while {\tt NNPDF}s give slightly larger values. PDF effects almost completely cancel between $\sigma_A$ and $\sigma_S$ in the asymmetries. We therefore do not display the uncertainties of each individual PDF set, but have checked that their effect on the asymmetries is very small. Statistical uncertainties from Monte-Carlo event generation on the cross section and the LO asymmetries per bin are negligible. For $A_E$ ($A_E^{\rm opt}$) at NLO, the absolute statistical uncertainties per bin are below $0.1\permil$ ($0.5\permil$) in all bins within the range $\pi/10 < \theta_j < 9\pi/10$.

\section{Parton shower effects}\label{sec:shower}
Beyond fixed-order QCD corrections, in this section we consider effects on the energy asymmetry from multi-parton emissions described by the parton shower. Detailed analyses of parton shower effects have been performed for the rapidity asymmetries $A_y^j$ and $A_{|y|}^j$ in $t\bar t + j$ production~\cite{Alioli:2011as} and for $A_y$ in inclusive $t\bar t$ production~\cite{Skands:2012mm,Hoeche:2013mua}. We have performed a basic parton shower analysis of the energy asymmetry $A_E^{\rm opt}$ in $t\bar t + j$ production.

The matching of the hard subprocesses to the parton shower has been performed by means of the {\tt MC@NLO} framework as implemented in {\tt MadGraph5$\_$aMC@NLO}~\cite{Alwall:2014hca}. Partons from the obtained events have been showered using the Monte-Carlo program {\tt PYTHIA} version 6.428~\cite{Sjostrand:2006za}. The entire procedure has been carried out within {\tt MadGraph5$\_$aMC@NLO}, which ensures consistent subtraction of divergences for NLO matching and allows us to compare LO and NLO predictions matched to the parton shower (which we will call LO+PS and NLO+PS, respectively) in the same framework. As in the fixed-order analysis, top-quarks have been treated as stable particles. Final-state radiation from top-quarks has been included, but gives a small effect compared with radiation from light partons. Hadronization or underlying event effects are not taken into account. All other shower parameters are kept at their default values. Jet reconstruction and further settings are the same as for the fixed-order analysis described in Section~\ref{sec:lhc}.

When generating events, we have applied a cut on the parton $p$ with the highest transverse momentum at generation level, $p_T^p > 70\,\text{GeV}$. This cut prevents loosing too many events when applying the cut on the hardest jet after showering and jet clustering, $p_T^j > 100\,\text{GeV}$, as in our fixed-order analysis. It is important to choose $p_T^p$ significantly smaller than $p_T^j$, since the parton shower can enhance the transverse momentum of $p$ so that it passes the analysis cuts. We have checked that our results do not change if we lower $p_T^p$ even further, which confirms that our generation cut has been set low enough to ensure physical results. Based on this framework, we have generated 130 million events. This large number of events is necessary to bring statistical uncertainties from Monte-Carlo event generation down to a moderate level.

\begin{figure}[t]
\centering
\vspace*{-0.5cm}
\resizebox{1.0\linewidth}{!}{
\begin{tabular}{cc}
\includegraphics[scale=0.59]{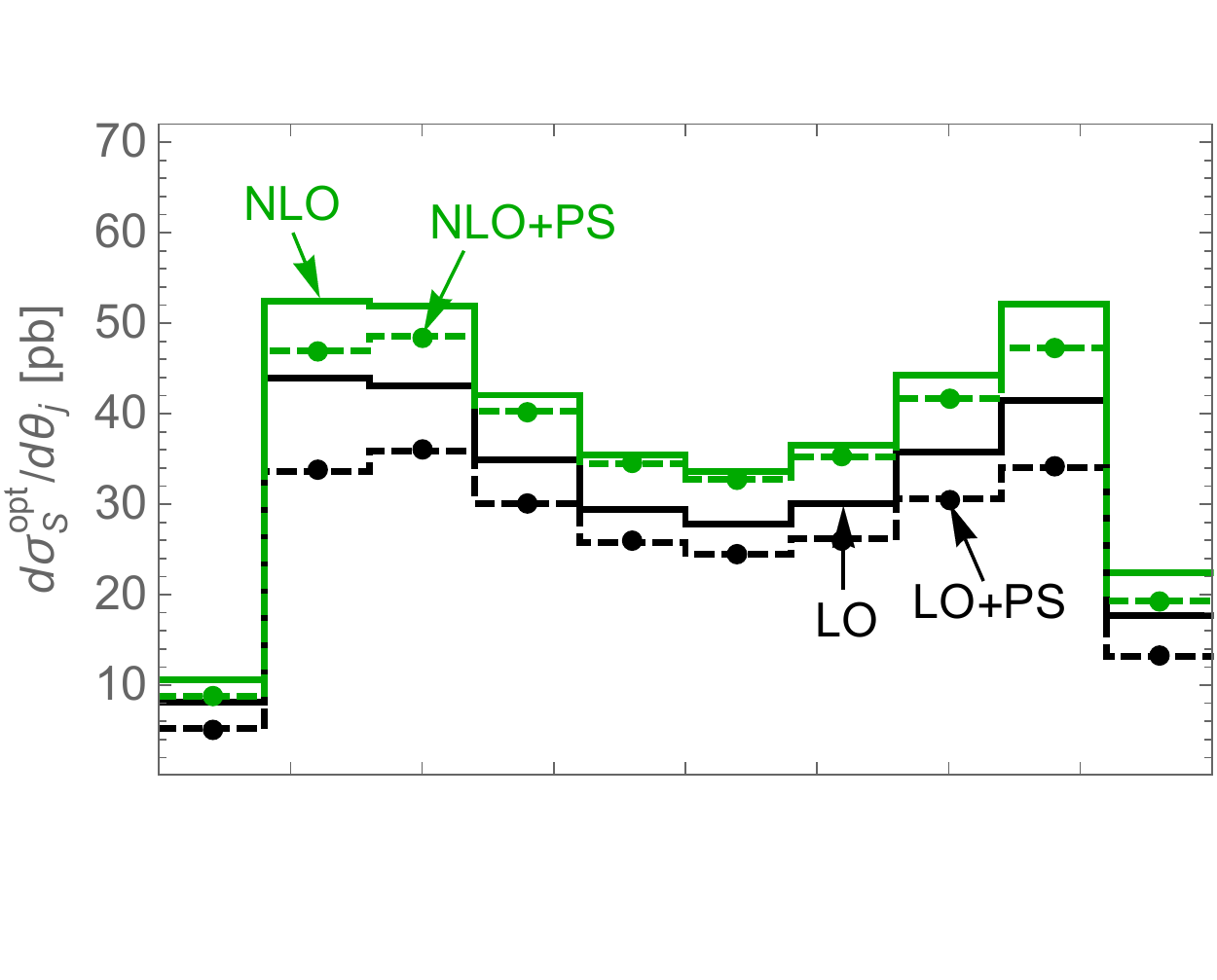}\hspace*{0.5cm} & \raisebox{-0.1cm}{\includegraphics[scale=0.607]{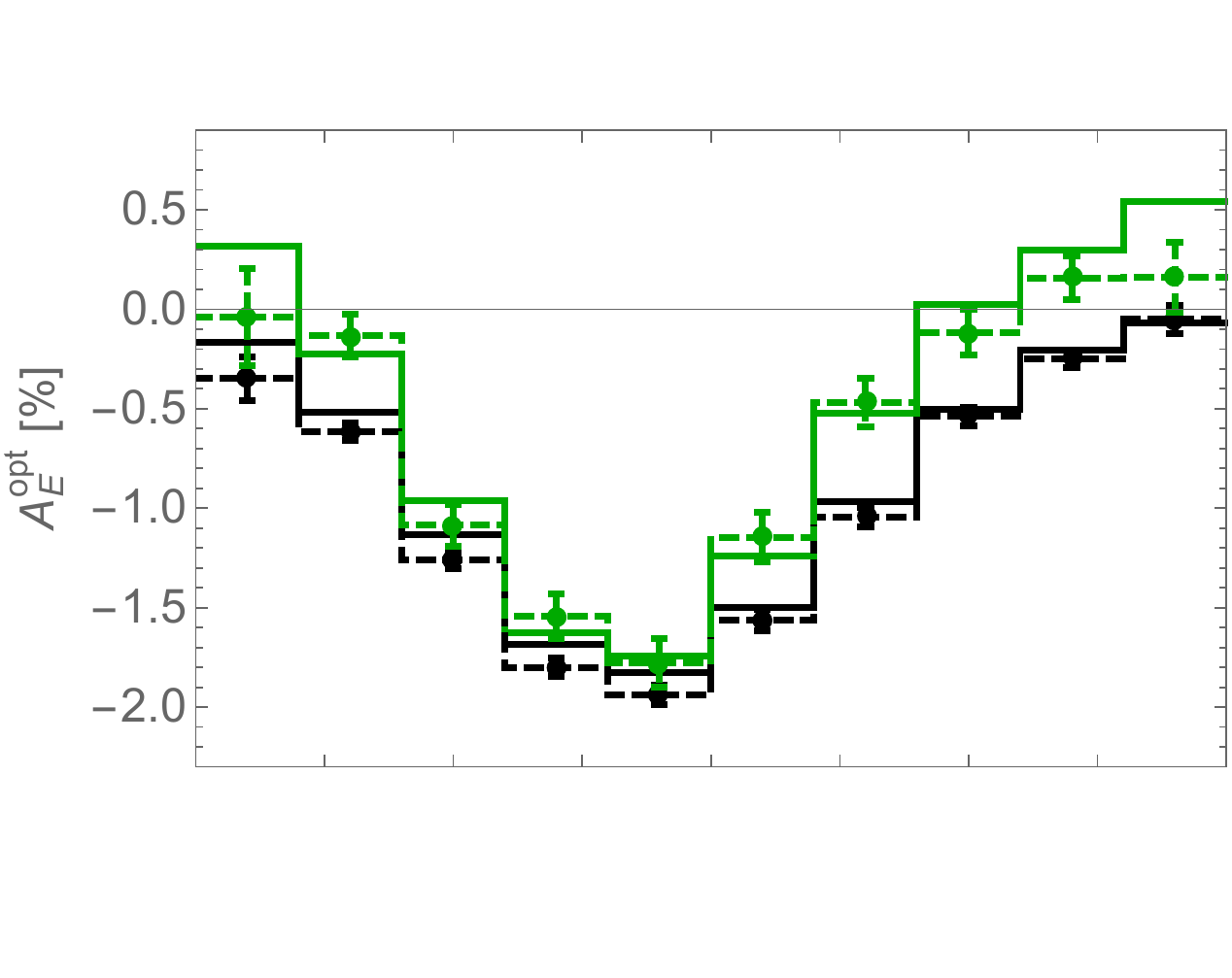}}\\[-1.95cm]
\hspace*{-0.07cm}\raisebox{0.1cm}{\includegraphics[scale=0.607]{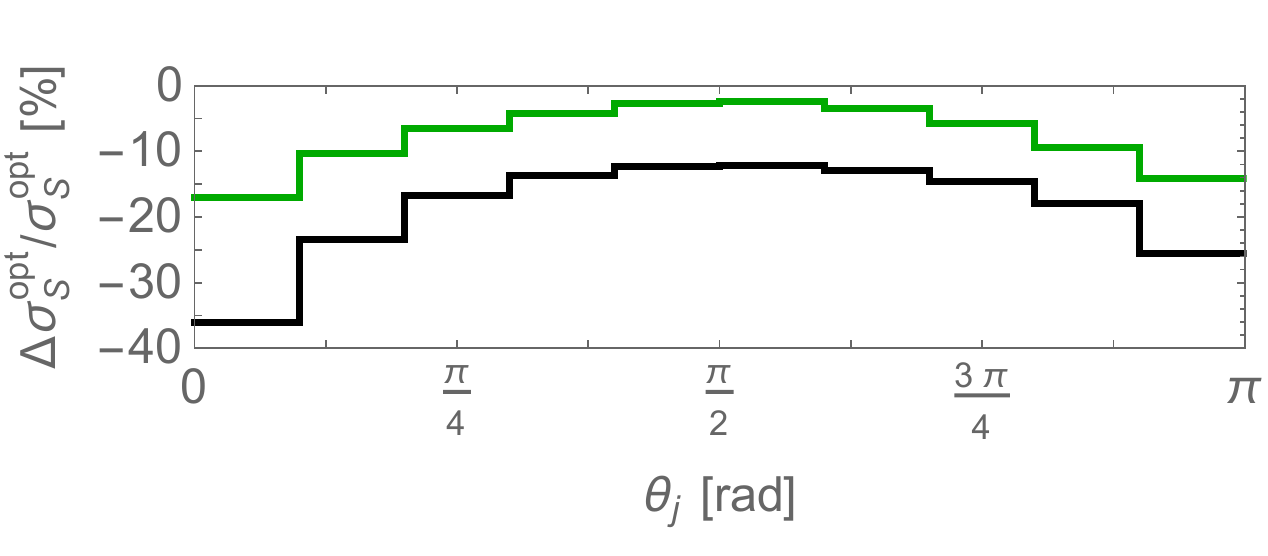}}\hspace*{0.55cm} & \hspace*{0.157cm}\includegraphics[scale=0.6213]{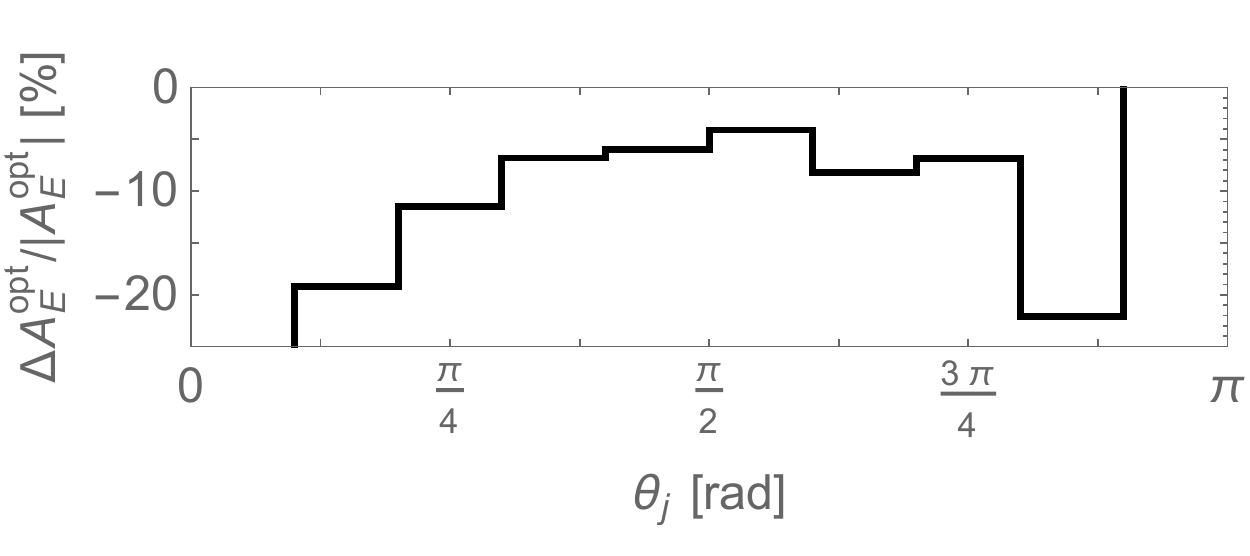}\\
\end{tabular}}\vspace*{-0.3cm}
\caption{Parton shower effects on the optimized energy asymmetry (right) and norma\-li\-zation (left) at LO(+PS) (black) and NLO(+PS) (green). Results with (without) parton shower effects are shown as dashed (plain) lines. Error bars indicate statistical uncertainties from Monte-Carlo event generation. Smaller panels show the relative effect of the parton shower. Cuts of $p_T^j > 100\,\text{GeV}$ and $|y_j| < 2.5$ have been applied.}
\label{fig:ps}
\end{figure}

Our results for the optimized energy asymmetry and the corresponding cross section are given in terms of jet distributions in Figure~\ref{fig:ps}. Fixed-order LO and NLO predictions are presented as plain lines, while LO+PS and NLO+PS predictions including parton-shower effects are shown as dashed lines. Error bars indicate the statistical uncertainty from Monte-Carlo event generation. The smaller panels show the relative change of the observable $O$ by the parton shower, according to $\Delta O/O = (O_{\rm (N)LO+PS}-O_{\rm (N)LO})/O_{\rm (N)LO}$. Since the uncertainty on the asymmetry at NLO+PS is sizeable even with the large data set, we have doubled the bin size in our distributions and refrained from displaying the relative change of the asymmetry.

From the left panel in Figure~\ref{fig:ps}, we see that the parton shower reduces the cross section compared to the fixed-order prediction. The effect is most pronounced near the endpoints of the distribution around $\theta_j = 0,\pi$, where the cross section features a collinear enhancement. Therefore, a change of the jet momentum by the parton shower can affect the rate sig\-ni\-fi\-cantly. This effect is reduced at NLO, where the first emission of an additional parton is included in the hard matrix element. At NLO+PS, the parton shower reduces $\sigma_S^{\rm opt}$ by about $-15\%$ near the endpoints of the distribution and by $-3\%$ in the central region, compared with the fixed-order NLO prediction. Notice that for $y_{t\bar t j}>0$ and $\theta_j \approx 0$ the jet is more likely to be emitted collinearly with the beam axis than for $y_{t\bar t j}>0$ and $\theta_j \approx \pi$. Parton-shower effects are therefore largest near $\theta_j = 0$.

A similar behavior can be observed for the optimized energy asymmetry, shown in the right panel of Figure~\ref{fig:ps}. At LO+PS, parton-shower corrections are largest near $\theta_j = 0$ and slightly deepen the minimum of the asymmetry. At NLO+PS, a moderate reduction of the asymmetry is observed near the endpoints of the distribution. The relative smallness of parton-shower effects is likely due to the strong cut on the jet transverse momentum, $p_T^j > 100\,\text{GeV}$. It reduces the sensitivity of the observable to phase-space regions with soft and collinear jet emission, where parton-shower effects can be important.

\section{Kinematic cuts and predictions for the LHC}\label{sec:kinematics}
After having determined the properties and uncertainties, in this section we make numerical predictions of the optimized energy asymmetry in $pp\to t\bar t j$ at NLO QCD for the LHC at $\sqrt{s}=13\,\text{TeV}$.\footnote{We do not include parton-shower effects in this section.} Our goal is to provide an observable that is well-suited for a measurement during run II. To this end, we investigate the effect of kinematic cuts that further enhance the asymmetry over its charge-symmetric background.

Two kinematic variables are especially useful to enhance the energy asymmetry. These are the top-antitop energy difference in the $t\bar t j$ frame, $\Delta E$, and the boost of the final state in the laboratory frame, $y_{t\bar t j}$. The dependence of the asymmetry on these variables has been studied in detail in Reference~\cite{Berge:2013xsa} for the energy asymmetry $A_E$. The main features apply as well to the optimized energy asymmetry $A_E^{\rm opt}$: The partonic energy asymmetry grows monotonically with its defining variable $\Delta E$ (just as the rapidity asymmetry $A_{y}^{(j)}$ grows with the top-antitop rapidity difference $\Delta y$). A cut on $y_{t\bar t j}$ helps to suppress partonic $gg$ background to the hadronic observable by selecting the boosted $qg$ initial states. Since the magnitude of the (optimized) energy asymmetry is not sensitive to the sign of $\Delta E$ and $y_{t\bar t j}$, we apply cuts on their absolute values.

\begin{figure}
\centering
\vspace*{-0.5cm}
\begin{tabular}{cc}
\includegraphics[scale=0.605]{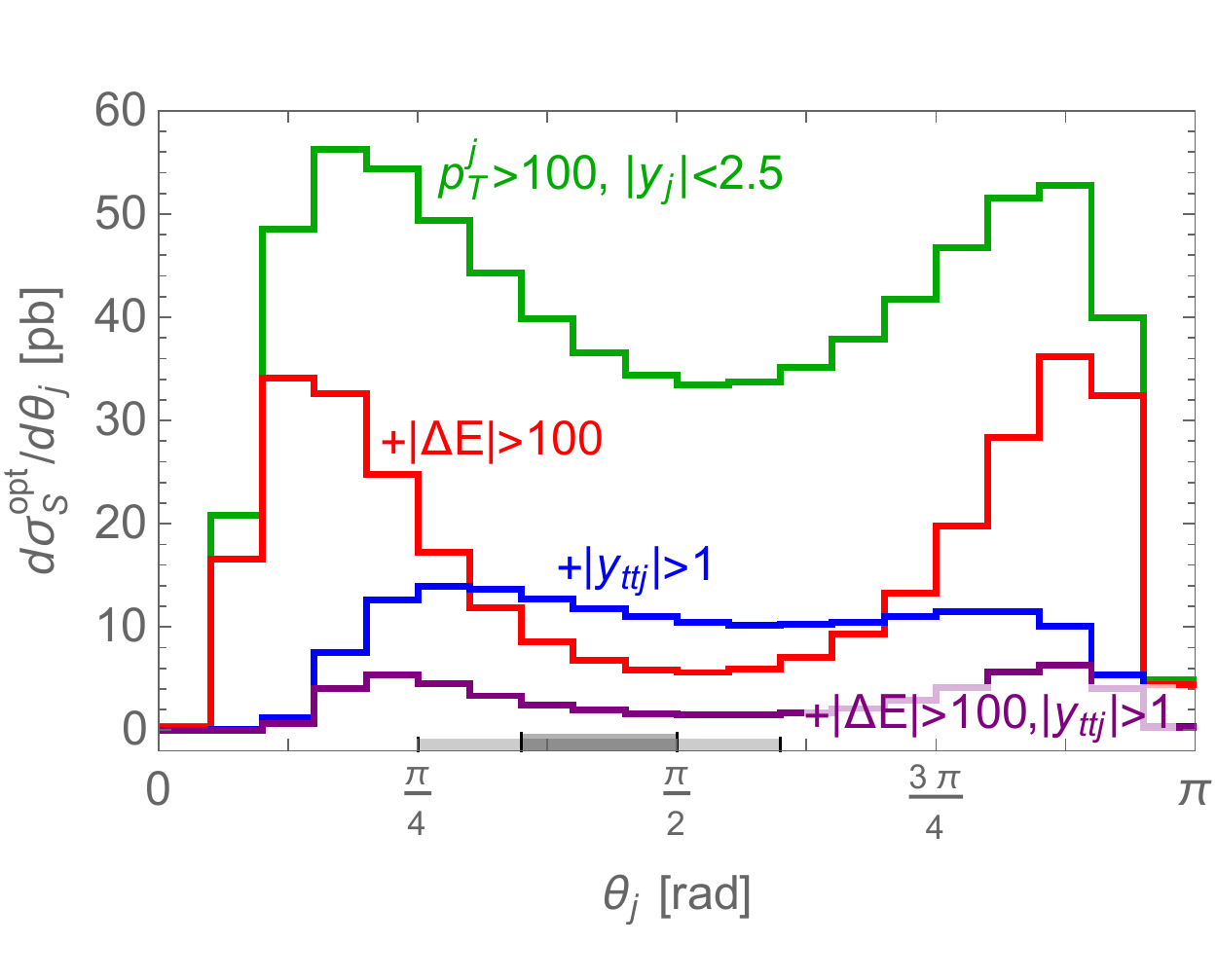}\hspace*{0.3cm} & \includegraphics[scale=0.6]{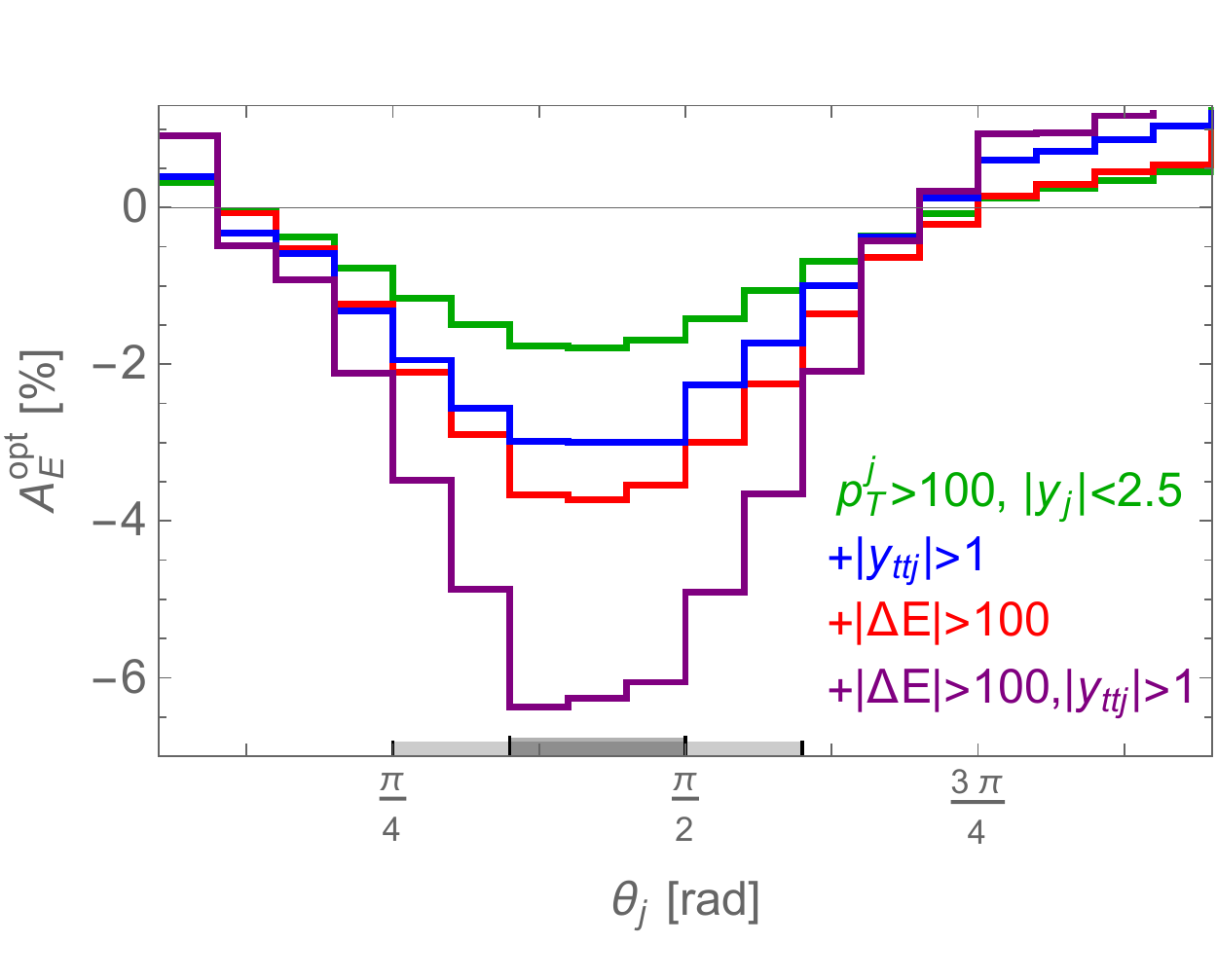}\\
\end{tabular}\vspace*{-0.3cm}
\caption{Kinematic cuts on the optimized energy asymmetry (right panel) and its nor\-ma\-li\-zation (left panel) at NLO QCD. The cuts $p_T^j > 100\,\text{GeV}$, $|y_j| < 2.5$ have been applied everywhere. Additional cuts on $|\Delta E|\,[\text{GeV}]$ and $|y_{t\bar t j}|$ are indicated in the figures. Gray bars mark the integration range of $\theta_j$ in Table~\ref{tab:predictions}.}
\label{fig:kin-cuts}
\end{figure}

The effect of these cuts on the jet distribution of the optimized energy asymmetry and its normalization is displayed in Figure~\ref{fig:kin-cuts}. The cut on $|\Delta E|$ is very efficient in enhancing the asymmetry (red curves). However, this comes at the price of sacrificing a large portion of the cross section in the central-jet region around $\theta_j\approx \pi/2$, where the asymmetry is largest and most precisely predicted. A lower cut on the boost $|y_{t\bar t j}|$ (blue curves), in turn, reduces the gluon-gluon background, ergo the cross section in the collinear regime $\theta_j\approx 0,\pi$. By combining both cuts we can find an optimal balance of enhancing the absolute asymmetry and preserving a sizeable data set. In our exemplary set of combined cuts, $|\Delta E| > 100\,\text{GeV}$ and $|y_{t\bar t j}| > 1$, the optimized energy asymmetry reaches $A_E^{\rm opt}\approx -6\%$ in its minimum (purple curves).

\begin{table}[!tb]
\centering
\renewcommand{\arraystretch}{1.5}
\begin{tabular}{|K{5.9cm}"K{4.4cm}|K{4.4cm}|}
\hline
$0 < \theta_j < \pi$ / $-\infty < \hat{y}_j < +\infty$ & no cut on $y_{t\bar t j}$ & $|y_{t\bar t j}| > 1$\\
\thickhline
no cut on $\Delta E$ & $\{120^{\,+\ 8}_{\,-15}, -0.6^{\,+0.2}_{\,-0.1}\}$ & $\{28^{\,+2}_{\,-4}, -1.2^{\,+0.3}_{\,-0.2}\} $ \\
\hline
$|\Delta E| > 100$ & $\{50^{\,+3}_{\,-6},-0.6^{\,+0.4}_{\,-0.2}\}$ & $\{8.5^{\,+0.5}_{\,-1.1},-1.5^{\,+0.7}_{\,-0.3}\}$\\
\hline
$|\Delta E| > 150$ & $\{32^{\,+2}_{\,-4},-0.4^{\,+0.5}_{\,-0.2}\}$ & $\{4.5^{\,+0.2}_{\,-0.6},-1.2^{\,+0.9}_{\,-0.4}\}$ \\
\hline
\end{tabular}\vspace*{0.4cm}
\begin{tabular}{|K{5.9cm}"K{4.4cm}|K{4.4cm}|}
\hline
$\frac{\pi}{4} < \theta_j < \frac{3\pi}{5}$ / $-0.32 < \hat{y}_j < 0.88$ & no cut on $y_{t\bar t j}$ & $|y_{t\bar t j}| > 1$\\
\thickhline
no cut on $\Delta E$ & $\{43^{\,+3}_{\,-5},-1.47^{\,+0.06}_{\,-0.05}\}$ & $\{13.2^{\,+0.8}_{\,-1.6},-2.51^{\,+0.09}_{\,-0.08}\}$\\
\hline
$|\Delta E| > 100$ & $\{9.7^{\,+0.2}_{\,-0.9},-2.88^{\,+0.08}_{\,-0.05}\}$ & $\{2.64^{\,+0.06}_{\,-0.25},-4.89^{\,+0.13}_{\,-0.09}\}$\\
\hline
$|\Delta E| > 150$ & $\{4.41^{\,+0.03}_{\,-0.74},-3.37^{\,+0.12}_{\,-0.07}\}$ & $\{1.105^{\,+0.001}_{\,-0.102},-6.5^{\,+0.1}_{\,-0.2}\}$ \\
\hline
\end{tabular}\vspace*{0.4cm}
\begin{tabular}{|K{5.9cm}"K{4.4cm}|K{4.4cm}|}
\hline
$\frac{7\pi}{20} < \theta_j < \frac{\pi}{2}$ / $0 < \hat{y}_j < 0.49$ & no cut on $y_{t\bar t j}$ & $|y_{t\bar t j}| > 1$\\
\thickhline
no cut on $\Delta E$ & $\{17^{\,+1}_{\,-2},-1.75^{\,+0.03}_{\,-0.03}\}$ & $\{5.6^{\,+0.4}_{\,-0.7},-2.99^{\,+0.03}_{\,-0.05}\}$\\
\hline
$|\Delta E| > 100$ & $\{3.34^{\,+0.01}_{\,-0.39}, -3.65^{\,+0.04}_{\,-0.19}\}$ & $\{0.94^{\,+0.01}_{\,-0.08},-6.25^{\,+0.07}_{\,-0.32}\}$\\
\hline
$|\Delta E| > 150$ & $\{1.46^{\,+0.02}_{\,-0.31},-4.28^{\,+0.04}_{\,-0.30}\}$ & $\{0.377^{\,+0.002}_{\,-0.061},-7.21^{\,+0.07}_{\,-0.42}\}$\\
\hline
\end{tabular}
\caption{Optimized cross section and energy asymmetry, $\{\sigma_S^{\rm opt}\,[\text{pb}],A_E^{\rm opt}\,[\%]\}$, with kinematic cuts on $|\Delta E|$ [GeV] and $|y_{t\bar t j}|$ at NLO QCD. The three tables correspond to different kinematic ranges of the jet angle $\theta_j$ (cf. Figure~\ref{fig:kin-cuts}). Uncertainties correspond to scale variations $\mu_R,\mu_F\in [0.5,2]\,m_t$.}
\label{tab:predictions}
\end{table}

Since the optimized energy asymmetry is largest near $\theta_j=\pi/2$, a further cut on the jet scattering angle should be applied. This cut can be expressed in terms of the jet rapidity in the $t\bar t j$ frame, $\hat{y}_j$, which can be measured as the difference between the jet rapidity and the final-state boost in the laboratory frame~\cite{Berge:2012rc},
\begin{equation}
\hat{y}_j\equiv\frac{1}{2}\ln\left(\frac{1+\cos\theta_j}{1-\cos\theta_j}\right) = y_j - y_{t\bar t j}.
\end{equation}
In Table~\ref{tab:predictions}, we present our predictions of the optimized energy asymmetry $A_E^{\rm opt}$ and the corresponding cross section $\sigma_S^{\rm opt}$ at NLO QCD for various combinations of kinematic cuts. In the three tables, the observables have been integrated over different ranges of the jet angle $\theta_j$ (or the corresponding jet rapidity $\hat{y}_j$). The errors correspond to scale variations $\mu_R,\mu_F\in [0.5,2]\,m_t$. Notice that the scale uncertainty is much reduced when cuts on $\theta_j$ are applied. This feature can be understood from Figure~\ref{fig:scale}, where the small panels show that the relative scale uncertainty is largest in the regions around $\theta_j = 0,\pi$. The improved precision of $A_E^{\rm opt}$ for $\theta_j\approx \pi/2$ is another strong argument for applying cuts on the jet direction. In our example with $|\Delta E| > 100\,\text{GeV}$ and $|y_{t\bar t j}| > 1$, the relative scale uncertainty of the asymmetry in the range $\frac{7\pi}{20} < \theta_j < \frac{\pi}{2}$ is no more than $5\%$.

\begin{figure}[t]
\centering
\includegraphics[scale=0.65]{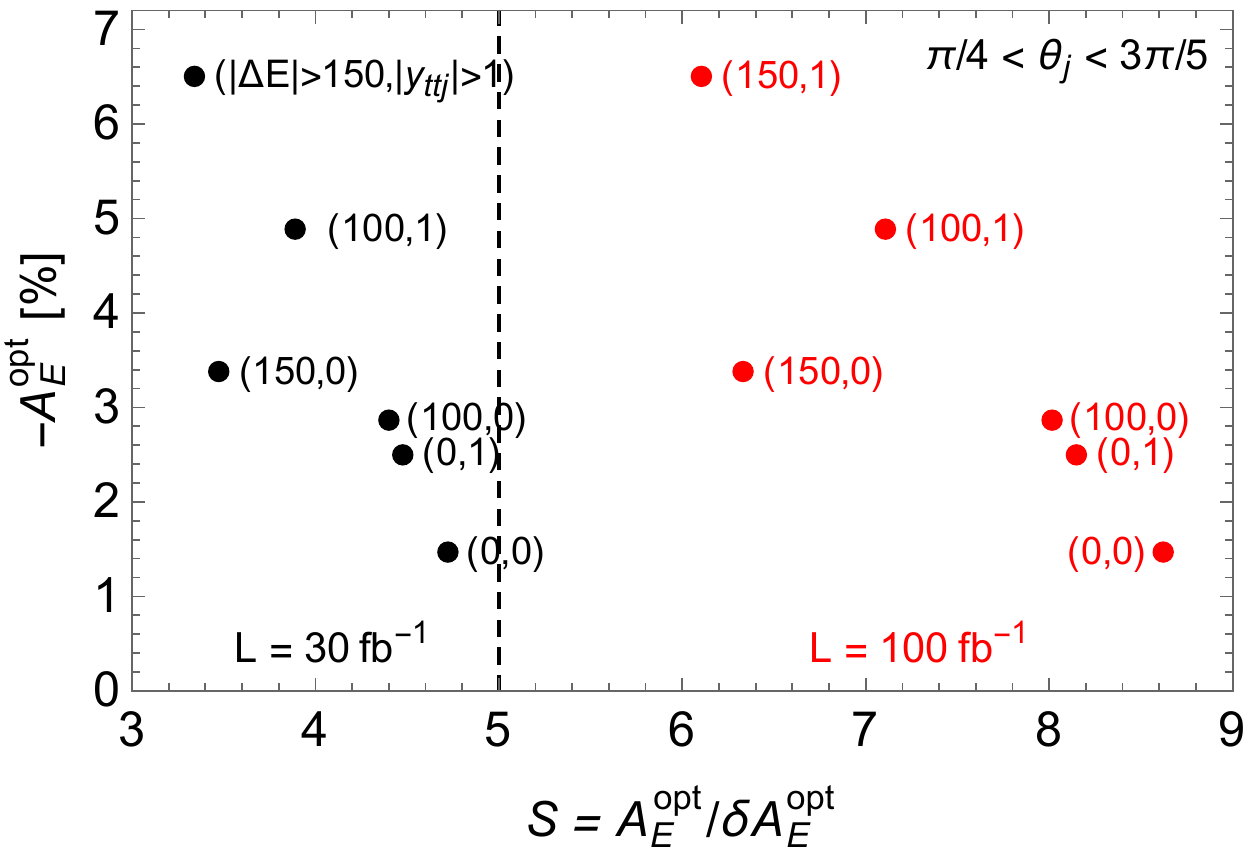}
\caption{Expected statistical significance $\mathcal{S}$ of the optimized energy asymmetry at the 13--TeV LHC during run II. Shown are the NLO predictions from Table~\ref{tab:predictions} for $\pi/4 < \theta_j < 3\pi/5$ for kinematic cuts $(|\Delta E|>a\,\text{GeV},|y_{t\bar t j}|>b)\to (a,b)$ and luminosities $\mathcal{L}=30\,\text{fb}^{-1}$ (black) and $\mathcal{L}=100\,\text{fb}^{-1}$ (red). An experimental efficiency of $\epsilon = 8\,\%$ is assumed. See text for details.}
\label{fig:stat-sig}
\end{figure}

To estimate the discovery prospects during run II, we have investigated the statistical significance of the optimized energy asymmetry. For a data set of $N$ Poisson-distributed events, the statistical uncertainty of the energy asymmetry is approximately given by $\delta A_E^{\rm opt}\approx 1/\sqrt{N}$. The number of $t\bar tj$ events is given by $N=\sigma_S^{\rm opt}\times \mathcal{L}\times \epsilon$, with the integrated detector luminosity $\mathcal{L}$ and the experimental efficiency $\epsilon$. The experimental efficiency denotes the combination of detector acceptance, selection efficiency and reconstruction to the parton level. In Figure~\ref{fig:stat-sig}, we show the statistical significance $\mathcal{S}=A_E^{\rm opt}/\delta A_E^{\rm opt}$ for NLO QCD predictions of the optimized energy asymmetry from Table~\ref{tab:predictions}. We make predictions for data sets of $\mathcal{L}=30\,\text{fb}^{-1}$ (black) and $\mathcal{L}=100\,\text{fb}^{-1}$ (red), corresponding with the luminosities recorded at the 13--TeV LHC until fall 2016 and expected during the complete run II, respectively. We assume an experimental efficiency of $\epsilon = 8\,\%$, roughly corresponding with the efficiency in current measurements of jet-associated top-pair production at 13~TeV.

From Figure~\ref{fig:stat-sig}, it is apparent that a sizeable energy asymmetry can be observed with a statistical significance of $\mathcal{S}\gtrsim 4\,(6)$ already now (by the end of run II). As expected, stronger kinematic cuts reduce the statistical significance. The ideal choice of cuts will ultimately depend on the size of remnant systematic uncertainties. In case of sizeable systematics, a larger asymmetry might be preferred at the expense of reducing the statistical significance. Notice, furthermore, that strong cuts might enhance the impact of higher-order QCD corrections on the specific phase-space region. In Section~\ref{sec:shower}, we have shown that the jet momentum cut does not lead to enhanced higher-order corrections from the parton shower. A similar dedicated analysis for the kinematic cuts described in this section would be valuable.

\section{Conclusions and outlook}\label{sec:conclusions}
In this work, we have advocated the energy asymmetry as an observable of the charge asymmetry in jet-associated top-antitop production. Our main goal was to answer the question if the energy asymmetry is stable under QCD corrections. To this end, we have calculated fixed-order NLO QCD contributions to the energy asymmetry. Corrections are moderate and most pronounced for jet emission collinear to the incoming partons (see Figure~\ref{fig:fo-lo-nlo}). In this region, new virtual NLO contributions shift the asymmetry to positive values, compared to its negative LO prediction. In the region of jet emission perpendicular to the incoming partons, where the absolute asymmetry is largest, NLO corrections are of a few relative percent only.

To estimate the uncertainty of our predictions, we have analyzed their dependence on the factorization and renormalization scales. While the inclusion of NLO contributions reduces the scale dependence of the cross section, it enhances the scale uncertainty of the asymmetry near the endpoints of the jet distribution~(see Figure~\ref{fig:scale}). This is due to the fact that the renormalization scale dependence no longer cancels between the numerator and denominator of the asymmetry, once NLO contributions are taken into account. The uncertainty is of a few percent only for perpendicular jet emission, but reaches several tens of percent in the collinear region.

These observations suggest to apply a cut on the jet angle $\theta_j$ that removes events with collinear jet emission. The asymmetry can be further enhanced by applying kinematic cuts on the top-antitop energy difference, $\Delta E$, and the boost of the final state, $y_{t\bar t j}$ (Figure~\ref{fig:kin-cuts}). Our final results are shown in Table~\ref{tab:predictions}, where we make predictions for the optimized energy asymmetry at the LHC with $\sqrt{s}=13\,\text{TeV}$. For instance, with $\pi/4 < \theta_j < 3\pi/5$, $|\Delta E| > 150\,\text{GeV}$ and $|y_{t\bar t j}| > 1$, corresponding to a cross section of $\sigma_S^{\rm opt}= 1.1\,\text{pb}$, an asymmetry of $A_E^{\rm opt} = -6.5\%$ can be reached. With tighter cuts, the asymmetry can be enhanced even more, but at the cost of further reducing the data set. Experimentally, the sensitivity to the energy asymmetry around its minimum might be improved by using dedicated reconstruction techniques for boosted top-quarks, as it has recently been done for the rapidity asymmetry at the LHC~\cite{Aad:2015lgx}.

Beyond fixed-order QCD, we have investigated the effect of multi-parton emission on the energy asymmetry (see Figure~\ref{fig:ps}). Due to the strong cut on the jet's transverse momentum, the impact of the parton shower on the observables is rather small in the region of perpendicular jet emission. However, the shower affects the collinear region, where effects are generally expected to be larger. In this work, we have confined ourselves to a basic study of the parton shower. A more comprehensive analysis would be very welcome, in particular to investigate the impact of the parton shower in the presence of kinematic cuts.

Our predictions are intended to serve as a basis for a measurement of the top charge asymmetry at the LHC. The energy asymmetry is sizeable in the region around its kinematic minimum, which we expect to be experimentally accessible to good precision with run-II data. In particular, the energy asymmetry is larger than rapidity-based asymmetries at the 13-TeV LHC in a data set of similar size, despite the additional jet in the final state. This makes us confident that systematic uncertainties of the experimental analysis can be reduced to a level that facilitates a discovery during run II. Obviously, the theory prediction of the energy asymmetry can be further improved by investigating the effects of top-quark decay~\cite{Bevilacqua:2015qha}, by including higher-order QCD corrections, and by studying the impact of phase-space cuts through a full-fledged parton shower analysis. Electroweak corrections, which are known to be sizeable for the rapidity asymmetries~\cite{Hollik:2011ps}, could also lead to significant effects. The careful investigation of the rapidity asymmetries at the Tevatron and the LHC has shown that such improvements are possible with the joint effort and ability of our community.

\subsection*{Acknowledgements}
We wish to thank Simone Alioli and Manfred Kraus for contributing to this project at an early stage. We thank Efe Yazgan and Bugra Bilin for discussions of experimental aspects, Jan Winter for helpful comments on the manuscript, Marco Zaro for his support with {\tt aMC@NLO}, and Torben Schell for help with the computing cluster. SW acknowledges fun\-ding by the Carl Zeiss foundation through a {\it Junior-Stiftungsprofessur}. We also acknowledge support by the state of Baden--W\"urttemberg through bwHPC.

\end{document}